\documentclass[a4paper,fleqn,usenatbib]{mnras}

\usepackage{newtxtext,newtxmath}

\usepackage[T1]{fontenc}
\usepackage{ae,aecompl}

\usepackage{graphicx}	
\usepackage{amsmath}	
\usepackage{amssymb}	

\usepackage{multirow}
\usepackage{textcase}
\usepackage{soul}

\title[KIC\,3240411]{KIC\,3240411 -- the hottest known SPB star with the asymptotic g--mode period spacing}

\author[Szewczuk \& Daszy{\'n}ska-Daszkiewicz]{
Wojciech Szewczuk\thanks{E-mail: szewczuk@astro.uni.wroc.pl (WS)},
Jadwiga Daszy{\'n}ska-Daszkiewicz
\\
Astronomical Institute of the Wroc{\l}aw University, Kopernika 11, PL-51-622 Wroc{\l}aw, Poland
}

\date{Accepted XXX. Received YYY; in original form ZZZ}

\pubyear{2015}

\begin{document}
\label{firstpage}
\pagerange{\pageref{firstpage}--\pageref{lastpage}}
\maketitle

\begin{abstract}
We report the discovery of the hottest hybrid B-type pulsator, KIC\,3240411, that exhibits the period spacing
in the  low-frequency range. This pattern is associated with asymptotic properties of high-order gravity (g-) modes.
Our seismic modelling made simultaneously with the mode identification shows that
dipole axisymmetric modes best fit the observations.
Evolutionary models are computed with MESA code and pulsational models with the linear non-adiabatic code employing
the traditional approximation to include the effects of rotation.
The problem of mode excitation is discussed. We confirm that significant modification is indispensable
to explain  an instability of both pressure and gravity modes in the observed frequency ranges of KIC\,3240411.
\end{abstract}




\section{Introduction}

Slowly Pulsating B-type stars (SPB) were identified as a separate class of pulsating variables
by \citet{Waelkens1991}. Initially, the definition of this class included main-sequence stars
of spectral types between B3 and B9, which correspond to masses in the range 2.5--7 $\mathrm  M_{\sun}$.
They exhibit multiperiodic variations in the light and spectral lines with typical  periods of about 0.5 to 3 days.
 It is now well established that these variations are associated with heat-driven pulsations in high-order gravity (g) modes
\citep{WD_PM_AP1993,Gautschy1993}.

In the era of large multisite campaigns and, especially, satellite observations
this definition has become somewhat imprecise because high-order g modes have been detected
in main-sequence stars hotter than B3 spectral type, e.g., $\nu$ Eri \citep{2005MNRAS.360..619J}, 12 Lac \citep{2006MNRAS.365..327H},
$\gamma$ Peg \citep{2006A&A...448..697C}, and many early B-type stars observed by {\it Kepler} \citep{Balona2011}.

According to the theoretical predictions, the oscillation spectra
of SPB pulsators are very dense and in the asymptotic regime (high radial orders $n$)
one should expect regularities in period spacing. Unfortunately, ground-based observations show up
sparse oscillation spectra \citep{Waelkens1991, DeCat2002, DeCat2007}.
This view has changed tremendously with the advent of the space-borne observations.
Photometric data from the projects such as MOST, CoRoT, Kepler, and recently BRITE, revealed
huge multi-periodicity and diversity of pulsations of B--type stars. In particular, we learnt
that a lot of them, if not the vast majority, are hybrid pulsators of SPB/$\beta$ Cep or $\beta$ Cep/SPB type
\citep{Degroote2010N, Balona2011, McNamara2012, Papics2012,Papics2014, Papics2015, Papics2017,
2015MNRAS.451.1445B, 2016A&A...588A..55P, 2017A&A...603A..13K}.

\begin{figure*}
	\includegraphics[width=1.45\columnwidth, angle=270]{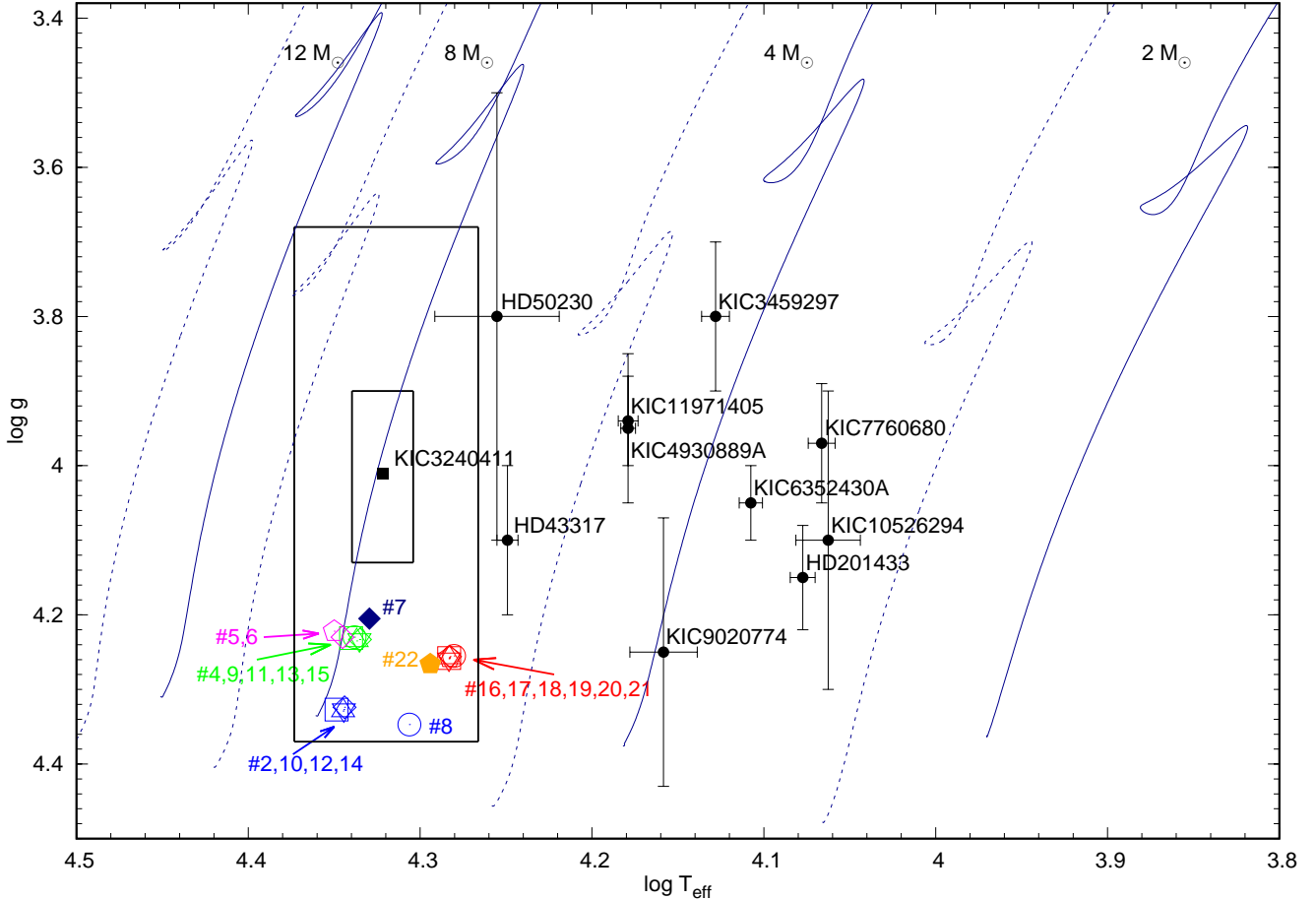}
    \caption{Currently known B-type pulsators exhibiting period spacing patterns in the Kiel diagram.
    For KIC\,3240411 we depicted the two error boxes;  $1\sigma$ and $3\,\sigma$. For other stars, only $1\,\sigma$ error bars is shown.
             Also evolutionary tracks for non-rotating models
             calculated with the OPLIB opacity tables assuming $X=0.738$, $Z=0.013$ (solid dark-blue lines) and $X=0.670$, $Z=0.006$
             (dashed dark-blue lines) are shown. Small value of overshooting from the convective core, $f_\mathrm{ov}=0.02$, was assumed .
             The big symbols with numbers will be explained later.
             The stellar parameters ($\log T_\mathrm{eff}$ and $\log g$) are taken from the literature (for the values and references see Table\,B1).
             Colored figures are available only in the electronic version.
             }
    \label{Kiel}
\end{figure*}

Moreover, 10 SPB stars with regular period spacing patterns were found
\citep{Degroote2010N, Papics2012, Papics2014, Papics2015, Papics2017, 2017A&A...603A..13K}.
The positions of these all stars on the Kiel diagram are shown in Fig.\,\ref{Kiel}
and their main period spacings in Fig.\,B1.
The parameters  of these stars  and  references are listed in Table\,B1.
However, one has to be cautious because not every period spacing can be associated with the asymptotic properties.
For example,  \citet{2014IAUS..301..109S} have shown that in the case of HD\,50230
the period spacing found by \citet{Degroote2010N}  is most probably accidental.
In turn, in the case of HD\,43317, \citet{2013A&A...559A..25S} did not succeed in reproducing any
of the two period spacings.

The period spacing depends on the spherical harmonic degree $\ell$, azimuthal order $m$,
and it is very sensitive to evolutionary changes \citep{WD_PM_AP1993}
as well as to the effects of rotation \citep{2017MNRAS.469...13S}.
These properties allow seismic modelling of such pulsators and constraining parameters of a model and theory.

As has been shown by \citet{2015MNRAS.453..277S}, seismic modelling of SPB stars is possible
even if no asymptotic pattern is observed in their oscillation spectra, provided an unambiguous mode identification
is doable and the effects of rotation are properly included.
This demands high-quality, long time-series photometric and/or spectroscopic observations.
This was the case for HD\,21071 and, to our knowledge, this is the only example of seismic modelling of the SPB star
without period spacing structure.
Of course, the ideal situation would be to have both period spacing and mode identification for some frequency peaks
from photometric and/or spectroscopic observables.

So far, detailed seismic modelling based on period spacing was performed for the two SPB stars: KIC\,10526294 and KIC\,7760680.
In the case of KIC\,10526294, \citet{Moravveji2015} found that
exponentially decaying prescription for core overshooting describes the seismic data better than a step function and
derived a value of $f_\mathrm{ov}$ between 0.017 and 0.018.
The authors also found the need for additional diffusive mixing in the radiative envelope
with the value of $\log D_\mathrm{mix}$ between 1.75 and 2.00 dex (with $D_\mathrm{mix}$
in  cm$^{2}$ s$^{-1}$). Moreover, \citet{2015ApJ...810...16T}
constrained rotation rate near the core-envelope boundary to $163\pm89$ nHz.
The authors concluded that the seismic data are consistent with a rigid rotation
but a profile with counter-rotation within the envelope has a statistical advantage  over constant rotation.
For KIC\,7760680 \citet{Moravveji2016} constrained the overshooting
parameter to $f_\mathrm{ov}=0.024 \pm 0.001$. Similarly as in the case of KIC\,10526294,
the authors found the need for  extra diffusive mixing in the radiative envelope
that they confined to $\log D_\mathrm{mix}=0.75 \pm 0.25$.
But they were unable to find model in which all theoretical modes associated with observed frequencies are unstable.
This was succeeded by \citet{2017EPJWC.16003012S} who applied appropriate modifications of the opacity data.

Here, we report the discovery of the hottest known SPB star exhibiting regular period spacing pattern found in the Kepler data.
In Section 2 we describe the star KIC 3240411 and in Section 3 its space photometry from Kepler and results of the frequency analysis.
Extensive seismic modelling is presented in Section 4. Conclusions and discussion close the paper.

\section{The star KIC\,3240411}

\begin{figure*}
	\includegraphics[width=\columnwidth, angle=270]{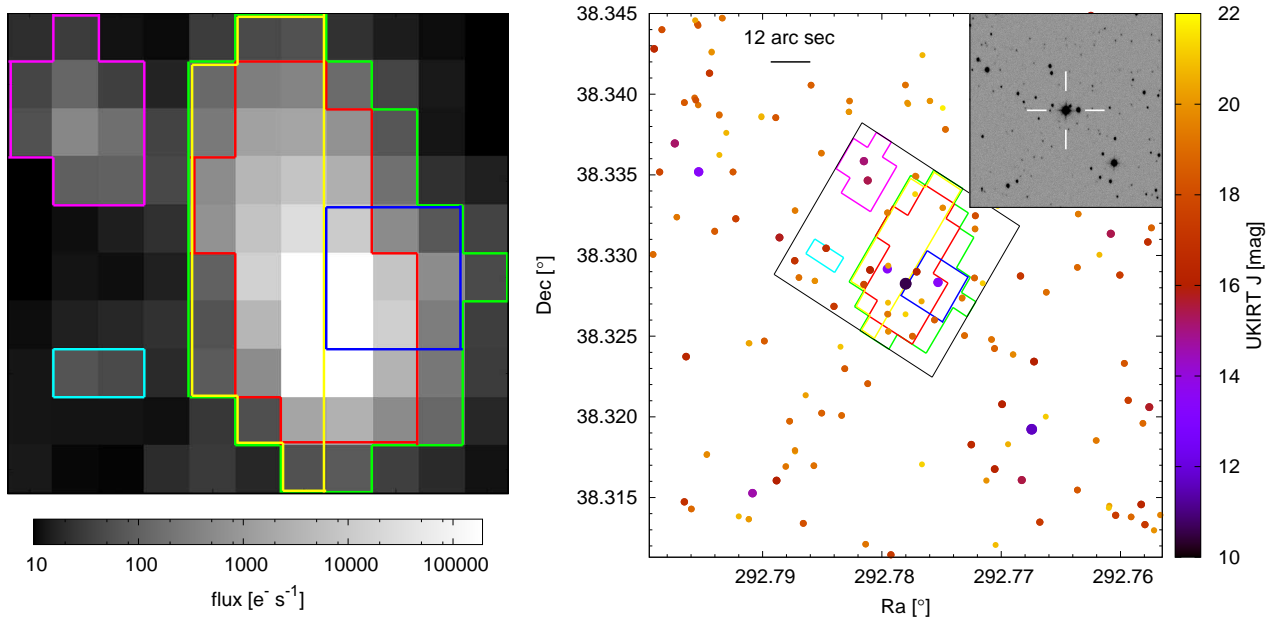}
    \caption{The left-hand panel: one of the middle frames from the target pixel file for Q5 dataset of KIC\,3240411.
           Different masks are shown used for extracting the aperture photometry: red colour -- an original mask, green -- a mask defined by us,
           magenta and cyan -- masks for the two possibly variable sources, yellow and blue -- submasks for checking differences in the
           Fourier amplitude spectrum within mask for our target (see the text for more explanation).
         The right-hand panel:  the positions of stars on the sky around KIC\,3240411 and their UKIRT J magnitudes coded by colours.
           The black square encompasses the area shown in the left-hand panel. The inset in the top-right corner contains the original UKIRT frame.
             }
    \label{fig_field}
\end{figure*}

\begin{figure}
	\includegraphics[width=1.2\columnwidth, angle=270]{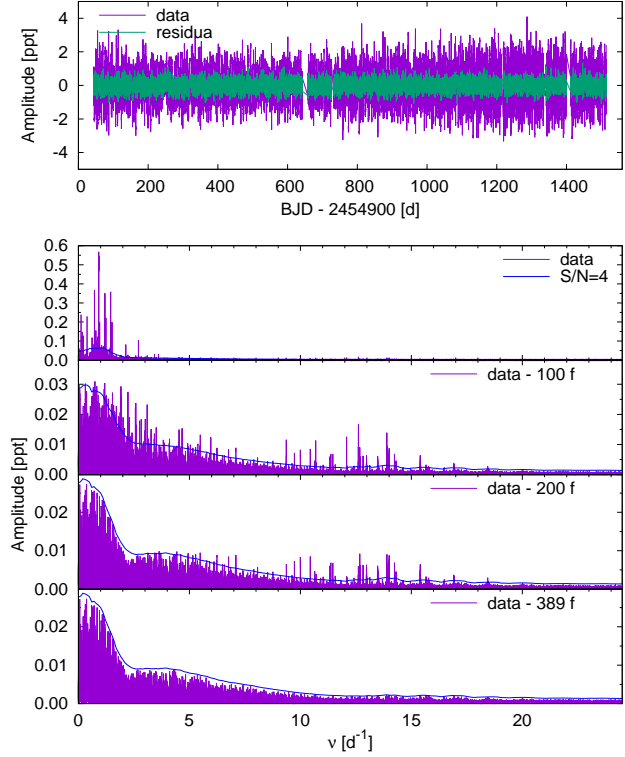}
    \caption{The light curve of KIC\,3240411 and the residua after prewhitening on all
      significant frequencies (the top panel). The lower panels show: the Fourier amplitude spectrum of the original data,
      the spectrum after prewhitening on 100 frequencies, the spectrum after prewhitening on 200 frequencies
      and the spectrum after prewhitening on all significant frequencies. The significance level of  $\mathrm{S/N}=4$ is marked as well.}
    \label{fig_periodogram}
\end{figure}

KIC\,3240411 is the early-type B2V star \citep{Lehmann2011} with the Kepler brightness Kp=10.24 mag.
Its basic stellar parameters, as determined by \citet{Lehmann2011} from spectroscopic observations, are:
the effective temperature, $T_\mathrm{eff}=20980^{+880}_{-840}$, surface gravity, $\log g=4.01^{+0.12}_{-0.11}$,
and the projected rotation velocity, $V_\mathrm{rot}\sin i=43 \pm 5$ km\,s$^{-1}$.
The authors found also helium overabundance, $Y=0.327$ and metallicity twice lower than for the Sun, $Z=0.006\pm 0.002$.
Using the $uvby\beta$ photometry, \citet{2011A&A...528A.148H} derived the similar values of the effective temperature and gravity:
$T_\mathrm{eff}=20900$\,K and $\log g=4.09$ with the errors of about 840\,K and 0.2 dex, respectively.
The position of the star on the Kiel diagram according to the parameters by \citet{Lehmann2011} is shown in Fig.\,\ref{Kiel}.
Two error boxes are depicted with uncertainty of 1$\sigma$ and 3$\sigma$.
The parallax of KIC\,3240411 is listed in the first Gaia data release \citep{2016A&A...595A...2G}
but its error exceeds the value more than four times, $\pi=0.08(37)$ mas.

KIC\,3240411 was classified as the SPB/$\beta$ Cep hybrid pulsator by \citet{Balona2011}
who found in total more than 100 frequency peaks based on the analysis of Q0-Q4 Kepler data.
This classification was confirmed by \citet{2015MNRAS.451.1445B} who extracted a large number of frequencies
with the most prominent ones below 2\,d$^{-1}$ and many very low-amplitude peaks up to 17\,d$^{-1}$.

\section{Kepler photometry}

KIC\,3240411 was observed by Kepler for 1470 days during the nominal satellite mission (Q0-Q17).
In the public domain, both the light curve and target pixel files are available.

Publicly available Kepler photometric light curves are extracted
from the pixels that lay inside the so-called optimal aperture (mask).
This aperture was selected to maximize the signal-to-noise ratio (S/N).
According to \citet{Papics2014} this is not the best choice for extracting
oscillation frequencies in B-type pulsators. Including additional
pixels in the mask provides better long-term stability of the light curves.
This statement was also confirmed in our tests.
Therefore we decided to construct light curves using customized masks.

In order to extract the light curve for KIC\,3240411, we used dedicated code called PyKE
\citep{2012ascl.soft08004S}. In the first step of our analysis we prepared masks for each quarter.
With some exceptions (e.g., possibility of the contribution of the flux from other source)
we selected all pixels with the flux above the level of $\sim 300$ e$^-$\,s$^{-1}$.
The chosen threshold is arbitrary but it gives a significantly smaller noise in the Fourier
periodogram for low frequencies.
An example mask for Q5 is presented in Fig.\,\ref{fig_field}.  The original mask is marked by red line whereas our mask is marked by
a green line. In addition we looked for variability for marked to sources (magenta and cyan lines).
 In the left panel, we show one of the frame from the middle of Q5 and in the right panel we show masks plotted on a wider field of view with nearby stars and their UKIRT J magnitudes \citep[]{2007MNRAS.379.1599L}. Using our customized masks we extracted light curve from target pixel files.
In the next step the obvious outliers were manually removed. Then we removed systematic trends using co-trending basis vectors.
An eye inspection of the light curves obtained with the use
of different sets of the basis vectors allowed to conclude that in the case of KIC\,324011 using vectors 1 and 2 is a reasonable
choice. Employing too many vectors could lead to overfitting the data, cleans off real astrophysical signals, or add spurious signals.
Finally, quarters were divided by a second-order polynomial fit and merged. The final light curve consisting of 65 766 data points
and converted to the units ppt
is shown in the top panel of Fig.\,\ref{fig_periodogram}. The small exception from this procedure was done for Q12 data.
In this case due to a strong dip of the flux we had to divide a quarter into two parts and detrend them separately.

\subsection{Frequency analysis}

\begin{figure}
	\includegraphics[width=1\columnwidth, angle=270]{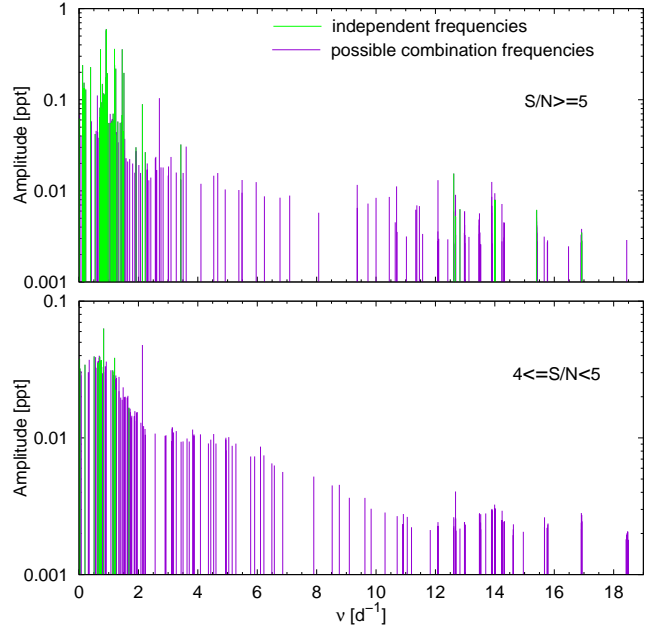}
    \caption{Frequencies and their amplitudes extracted from the light curve of KIC\,3240411.
             In the top panel  frequencies with $\mathrm{S/N}\ge5$ are shown, whereas in the bottom panel,
             frequencies with $4\le \mathrm{S/N} < 5$ are shown.
             The green vertical lines mark the frequencies that are definitely independent. The violet peaks indicate frequencies
             that may be combination frequencies using the pure mathematical condition.}
    \label{fig_freq_all}
\end{figure}

To extract the frequencies of variability we calculated the Fourier amplitude spectrum up to the Nyquist frequency ($\sim$24.5\,d$^{-1}$)
and identified the frequency with the highest amplitude.
Here, we applied discrete Fourier transform algorithm based on the concept of \citet{1985MNRAS.213..773K}.
Then, we fitted a sum of sine functions to the Kepler light curve in the form
\begin{equation}
S\left( t \right)= \sum^N_{i=1} A_i \sin \left[ 2\pi \left( \nu_i\,t+\phi_i  \right)  \right] +c,
\end{equation}
where $N$ is the number of sinusoidal components, $A_i,~\nu_i,~\phi_i$
are the amplitude, frequency, and phase of the $i$th component, respectively.
The offset $c$ ensures that $\int_T S(t)\, dt=0$, where  $T$ is a  time base.
In the next step, the original data were prewhitened on the all found frequencies and new Fourier
amplitude spectrum was calculated. The whole procedure was iteratively repeated
until our significance condition was met.

To estimate a significance level of extracted frequency peaks, usually one computes the signal-to-noise ($S/N$) ratio.
Most often the significance level of $S/N=4$ is assumed, which is the threshold found by \citet{1993spct.conf..106B}
for ground-based observations or by \citet{1997A&A...328..544K} for the Hubble Space Telescope observations.
However, \citet{2015MNRAS.448L..16B} showed that in the case of the Kepler data
this threshold should be higher, $S/N\approx5$.
Here we adopted $S/N=4$ as a tentative value and then increased the level up to $S/N=5$.
The noise $N$ was calculated as the mean amplitude in a 1\,d$^{-1}$ window around a given frequency peak.

In the bottom panel of Fig.\,\ref{fig_periodogram} we show four amplitude spectra (from top to bottom): 
for the original data, after prewhitening on 100, 200 and all frequencies.
As one can see below $\sim$2\,d$^{-1}$ there is a significant increase of the noise level which can suggest that there are still
some unresolved signals present in the data.

As a result of the prewhitening procedure we obtained 389 frequencies with $S/N\ge 4$ and 72
frequencies with $S/N\ge 5$.  The frequencies are listed in Table\,A1 in Appendix\,A.
The Rayleigh resolution of the full data set of KIC\,3240411 amounts to 0.00068 d$^{-1}$.
The last column of Table\,A1 lists frequencies closer than this resolution.
In Fig.\,\ref{fig_freq_all} we depicted the oscillation spectra for the two significance levels:
$4\le S/N < 5$ (the bottom panel) and $S/N \ge 5$ (the top panel).

The significant frequencies range up to $\sim 18.5 \,\mathrm d^{-1}$, but those with the highest amplitudes occupy the low-frequency range corresponding to pulsations in g modes. If we apply a simple mathematical condition, 319  of 389 detected frequencies
seem to be linear combinations, i.e.,
\begin{equation}
\nu_i=n\,\nu_j+m\,\nu_k,
\end{equation}
where $\nu_i$ is a child frequency, $\nu_j$, $\nu_k$ are parents frequencies
and $n,~m$ are integers in the range from $-10$ to $+10$.
These possible combinations are given in the penultimate column of Table\,A1 in Appendix\,A.
But taking into account the density of the oscillation spectrum, the fact that we need high coefficients in combinations and the fact that combinations mostly
do not result from the highest amplitude peaks, it is very probable that most of them are in fact independent frequencies.

\begin{figure}
	\includegraphics[width=0.935\columnwidth, angle=270]{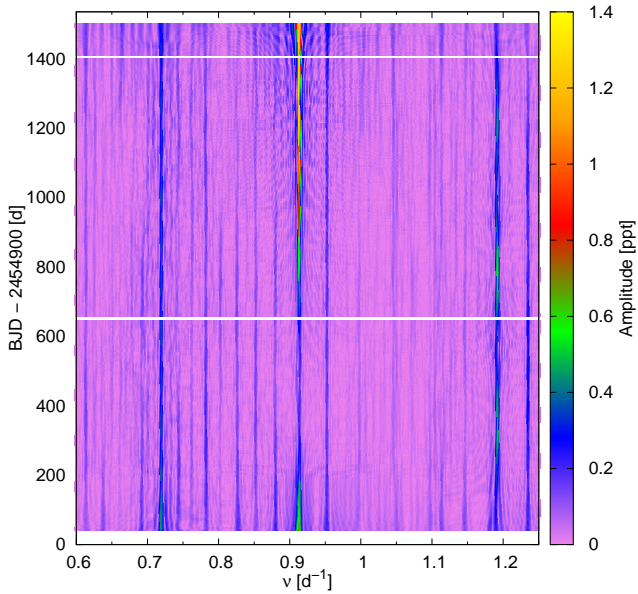}
    \caption{Time-depended Fourier amplitude spectrum calculated for the original data in the vicinity of the dominant frequency
    $\nu\approx 0.9156$ d$^{-1}$. The amplitude values are coded with colours.}
    \label{fourier_time_depend_MNRAS}
\end{figure}

Part of the variability, especially at high frequencies, may originate from the contamination by the fainter star, e.g., of $\delta$ Sct type.
Therefore we checked whether possible variable flux from stars within masks marked by magenta and cyan lines in Fig.\,\ref{fig_field}
contribute to the variability of KIC\,3240411. To this end, we constructed the light curves for these stars and followed the procedure
of the frequency analysis. We did not find any significant variability with frequency above $\sim0.1$\,d$^{-1}$.
The list of these frequencies is given in Table\,A2 in Appendix\,A.
We checked also the other possibility. Namely, there are two quite bright stars within our as well as the original mask of KIC\,3240411;
the one on the left and the other on the right. We calculated Fourier amplitude spectra of the light curves extracted from the pixels that lie
in the left half of our mask (pixels encounter by yellow line in Fig.\ref{fig_field}) and
from a few pixels on the right hand of the target star (pixels encountered by blue line in Fig.\ref{fig_field}).
Such cuts introduce significant trends, therefore we did not try to analyse
the merged data and focus only on Q5 data. Since in both sets of "cut" data as well as in the original data we found similar structures
in the Fourier amplitude spectrum at high frequencies, it suggests that all variability originates rather in KIC\,3240411.

\begin{figure}
	\includegraphics[width=0.935\columnwidth, angle=270]{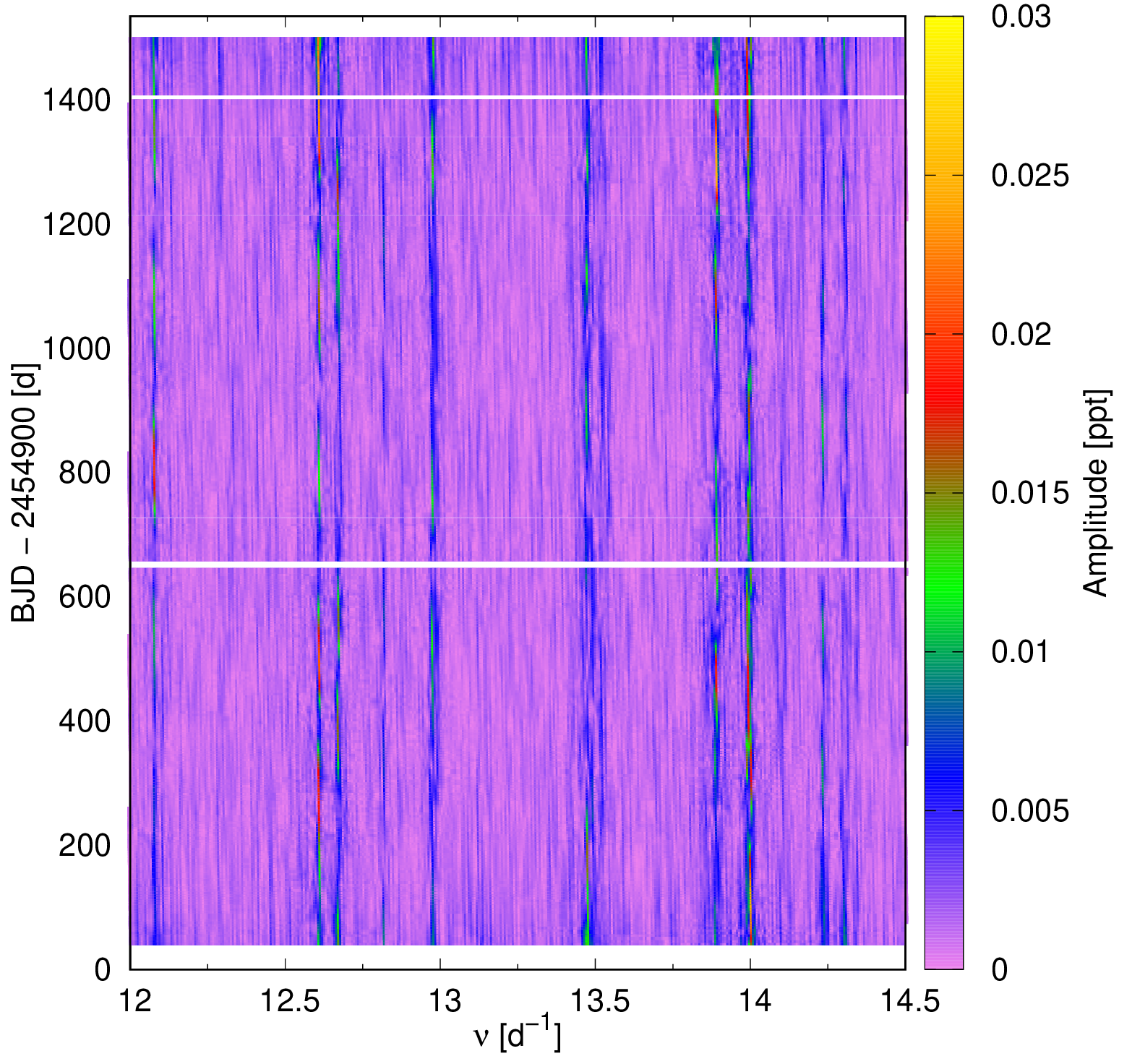}
    \caption{Same as in Fig.\,\ref{fourier_time_depend_MNRAS} but in the high-frequency domain and calculated
            for the data phrewithened on 149 frequencies.}
    \label{fourier_time_depend_MNRAS_high_nu}
\end{figure}

\begin{figure}
	\includegraphics[width=1.235\columnwidth, angle=270]{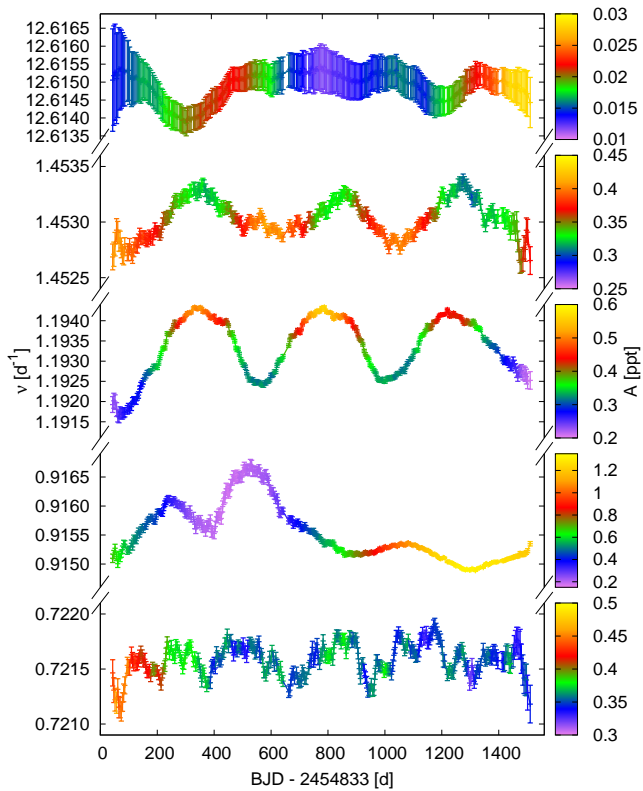}
    \caption{Changes of the frequencies (with the error bars) and their amplitudes with
      time. Four frequencies with the highest amplitudes
      (in the frequency range typical for g modes, the four bottom curves) and the one
      with much smaller amplitude but with the frequency value typical for p modes
      (the top curve) are shown. The colour scale for amplitudes are different for each frequency.}
    \label{freq_change}
\end{figure}

 In addition, we calculated time-depended Fourier amplitude spectra, i.e., the amplitude spectra for shorter segments of data.
We used different sizes of windows (from 50 to 300 d) and slid them along the time. The window shift was from about 0.2 to 6 d 
that corresponded to the number of observational points
from  10 to 300, respectively. We found quite significant changes of the frequencies and their amplitudes.
In Fig.\,\ref{fourier_time_depend_MNRAS}, we show time-depended Fourier amplitude spectrum calculated for the original
data in the vicinity of the dominant mode.  In this example, we used the window size of 300\,d shifted by 200 data points every time step.
As one can see the values of amplitudes, coded with colours, change, especially in case of the dominant frequency.
A larger spread of points in the second part of the light curve (see the top panel of Fig.\,\ref{fig_periodogram}) is caused mainly 
by the higher amplitude of the dominant frequency during the second part of observations.
Similarly, high frequencies change their amplitudes. The first  frequency above 10\,d$^{-1}$, i.e.,
$\nu_{150}=12.61497$\,d$^{-1}$, appears after
prewhitening of the original data on 149 frequencies. Therefore in order to emphasise changes in the high-frequency domain
we calculated time-depended Fourier amplitude spectrum for data after removing 149 frequencies (see Fig.\,\ref{fourier_time_depend_MNRAS_high_nu}).
A closer look reveals also that frequencies themselves are variable. In Fig.\,\ref{freq_change}, we show these changes for four low frequencies
and one high frequency. The frequencies and amplitudes shown in this figure were fitted to the data prewhitened on all frequencies with higher amplitudes
(with the exception of $\nu_1=0.915583$\,d$^{-1}$ which is frequency with the highest amplitude).

These changes depend slightly on the length of data subsets but the effect is not large.
On the one hand, the time base should be long enough to allow accurate determination of the  Fourier parameters.
On the other hand, longer the time base  larger the averaging of variations over time.
In Fig.\,\ref{freq_change}, we present the results of the Fourier analysis performed, as before,
for 300\,d data blocks shifted by 200 data point every time step.

The time-varying nature of frequencies on such short time scale may be caused by nonlinear effects
such as resonant couplings of pulsational modes \citep{1995NYASA.773....1B,1997A&A...321..159B}.  Amplitude modulation may result also
from beating of close frequency modes \citep{2017EPJWC.16003008B}.  As one can see for some modes there is a correlation between
the amplitude changes and frequency changes.
This can be caused by the beating phenomenon, nonlinearity, or mode coupling \citep[][and references therein]{2016MNRAS.460.1970B}.
Nevertheless, we can say that the frequency and amplitude variability is rather intrinsic  than  extrinsic as we did not find
any reliable signal of binarity in the light curve of KIC\,3240411 \citep{2014MNRAS.441.2515M}.

\subsection{Period spacing}

\begin{figure*}
	\includegraphics[width=1.4\columnwidth, angle=270]{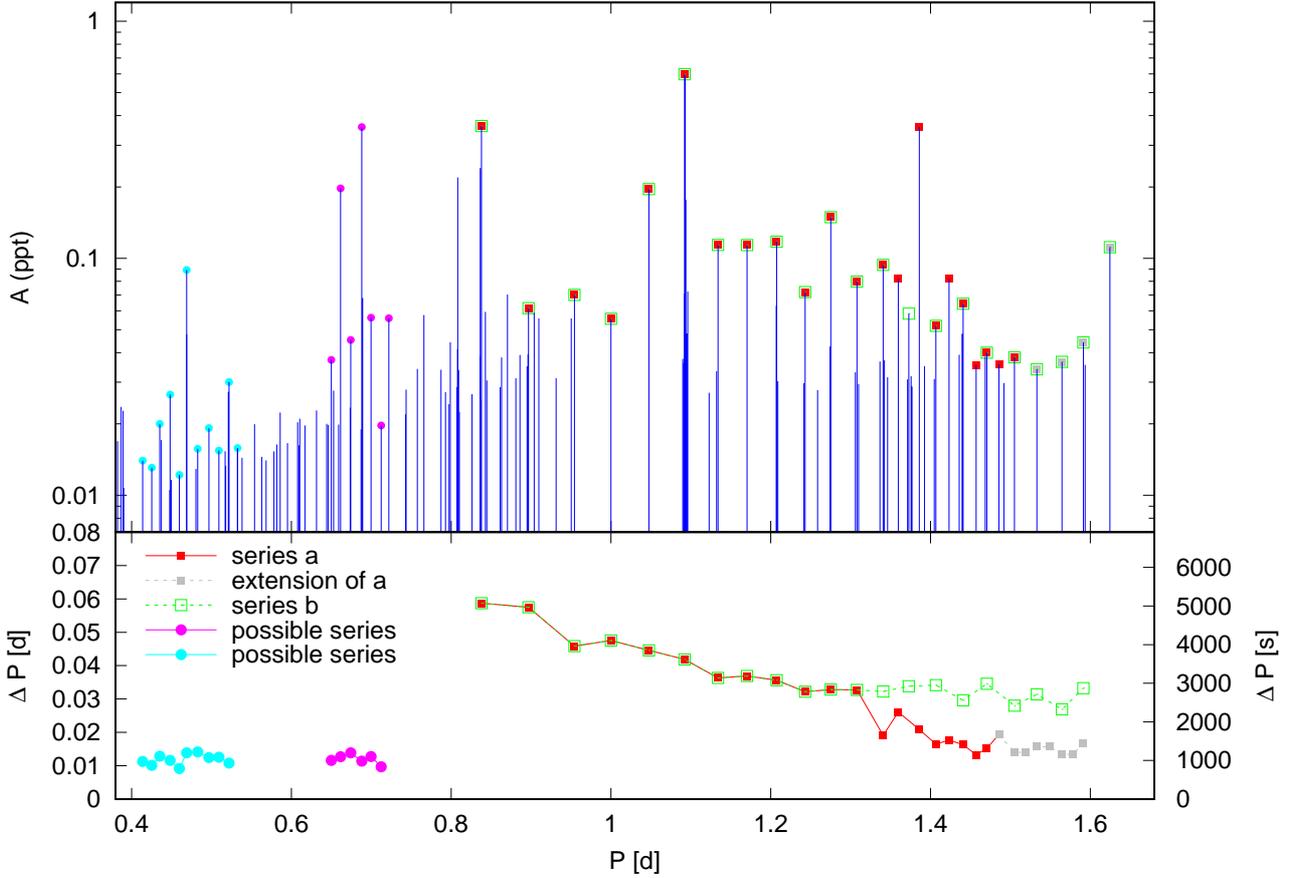}
         \caption{{\it Top panel:} The amplitude spectrum of KIC\,3240411 in the frequency range in which quasi-equidistant structures
         were found.  The frequencies that constitute series are marked by colour symbols. The series {\it a } is marked by red solid squares
         and  its possible extension by grey solid squares. The series {\it b} is marked by green open squares. Two additional series found in data
             are marked by cyan and magenta dots. {\it Bottom panel:} Period spacing of all identified sequences as a function of the period.}
    \label{per_spacings_KIC3240411}
\end{figure*}

\begin{table*}
	\centering
	\caption{The parameters of the two alternative (quasi) equidistant in period series. Frequencies, amplitudes,
	phases, periods, period spacing with their errors are provided here.
	In the penultimate column  the signal-to-noise ratios are listed and in the last  column  possible combinations are given.}
	\label{tab_freq_series}
	\begin{tabular}{rrrrrrrr}
	\hline
    ID&    $\nu$ [d$^{-1}$] & A [ppt] & $\phi$ [0--1] & $P$ [d] & $\Delta P$ [d] &  S/N & combinations\\
	\hline
	\multicolumn{8}{c}{a}\\
	\hline
   60 &    0.664458(19) &      0.0382(17) &      0.6320(70) &    1.504986(43) &                 &    5.1 & 2$\nu_{13}$-4$\nu_{17}$    \\
   65 &    0.673129(20) &      0.0356(17) &      0.3875(75) &    1.485599(44) &      0.01939(6) &    4.9 &     \\
   70 &    0.680081(21) &      0.0399(17) &      0.5501(68) &    1.470413(45) &      0.01519(6) &    4.8 & 2$\nu_{12}$-$\nu_{30}$    \\
   83 &    0.686283(21) &      0.0353(17) &      0.1888(75) &    1.457125(45) &      0.01329(6) &    4.7 & 8$\nu_{12}$-8$\nu_{66}$    \\
   36 &    0.694062(14) &      0.0643(17) &      0.7636(42) &    1.440793(29) &      0.01633(5) &    7.0 &     \\
   24 &    0.702686(11) &      0.0819(17) &      0.2899(32) &    1.423111(22) &      0.01768(4) &    9.1 &     \\
   37 &    0.710902(14) &      0.0518(17) &      0.0804(52) &    1.406664(28) &      0.01645(4) &    6.9 &     \\
    3 &    0.721591(04) &      0.3582(17) &      0.1980(07) &    1.385827(08) &      0.02084(3) &   27.5 &     \\
   28 &    0.735437(12) &      0.0817(17) &      0.1588(33) &    1.359736(22) &      0.02609(2) &    8.3 &     \\
   22 &    0.745916(09) &      0.0939(17) &      0.7233(29) &    1.340634(16) &      0.01910(3) &   10.3 &     \\
   25 &    0.764563(11) &      0.0797(17) &      0.1868(34) &    1.307937(19) &      0.03270(2) &    8.8 &     \\
   13 &    0.784238(07) &      0.1492(17) &      0.6323(18) &    1.275123(11) &      0.03281(2) &   14.5 &     \\
   33 &    0.804553(13) &      0.0720(17) &      0.6807(38) &    1.242926(20) &      0.03220(2) &    7.3 &     \\
   23 &    0.828305(10) &      0.1175(24) &      0.7586(33) &    1.207285(15) &      0.03564(2) &    9.9 &     \\
   18 &    0.854405(08) &      0.1136(17) &      0.3873(23) &    1.170405(11) &      0.03688(2) &   11.7 &     \\
   19 &    0.881783(08) &      0.1139(17) &      0.9447(23) &    1.134066(10) &      0.03634(2) &   11.6 &     \\
    1 &    0.915583(03) &      0.5981(19) &      0.5649(05) &    1.092200(04) &      0.04187(1) &   36.9 &     \\
   10 &    0.954553(06) &      0.1958(17) &      0.0933(14) &    1.047611(07) &      0.04459(1) &   18.3 &     \\
   40 &    0.999928(14) &      0.0556(17) &      0.4315(48) &    1.000072(14) &      0.04754(2) &    6.7 &     \\
   29 &    1.047990(13) &      0.0700(17) &      0.0747(38) &    0.954208(12) &      0.04586(2) &    7.6 & 5$\nu_{5}$-7$\nu_{24}$    \\
   31 &    1.115213(13) &      0.0616(17) &      0.0032(43) &    0.896690(10) &      0.05752(2) &    7.6 & -7$\nu_{14}$+2$\nu_{30}$    \\
    5 &    1.193358(04) &      0.3612(17) &      0.5842(07) &    0.837972(03) &      0.05872(1) &   30.6 &     \\
	\hline
	\multicolumn{8}{c}{b}\\
	\hline
   20 &    0.615542(08) &      0.1114(17) &      0.5774(24) &    1.624585(21) &                 &   12.1 & $\nu_{8}$+$\nu_{17}$    \\
   52 &    0.628407(18) &      0.0442(17) &      0.1445(61) &    1.591325(46) &      0.03326(5) &    5.6 & 6$\nu_{8}$-2$\nu_{18}$    \\ 	
   78 &    0.639212(21) &      0.0365(17) &      0.5658(73) &    1.564426(51) &      0.02690(7) &    4.8 & -10$\nu_{8}$+3$\nu_{11}$    \\
   87 &    0.652309(21) &      0.0340(17) &      0.0873(78) &    1.533016(49) &      0.03141(7) &    4.8 & $\nu_{32}$-4$\nu_{77}$    \\ 	
   60 &    0.664458(19) &      0.0382(17) &      0.6320(70) &    1.504986(43) &      0.02803(7) &    5.1 & 2$\nu_{13}$-4$\nu_{17}$    \\
   70 &    0.680081(21) &      0.0399(17) &      0.5501(68) &    1.470413(45) &      0.03457(6) &    4.8 & 2$\nu_{12}$-$\nu_{30}$    \\
   36 &    0.694062(14) &      0.0643(17) &      0.7636(42) &    1.440793(29) &      0.02962(5) &    7.0 &     \\
   37 &    0.710902(14) &      0.0518(17) &      0.0804(52) &    1.406664(28) &      0.03413(4) &    6.9 &     \\
   34 &    0.728408(14) &      0.0584(17) &      0.2098(46) &    1.372857(26) &      0.03381(4) &    7.1 & 2$\nu_{7}$-9$\nu_{15}$    \\
   22 &    0.745916(09) &      0.0939(17) &      0.7233(29) &    1.340634(16) &      0.03222(3) &   10.3 &     \\
   25 &    0.764563(11) &      0.0797(17) &      0.1868(34) &    1.307937(19) &      0.03270(2) &    8.8 &     \\
   13 &    0.784238(07) &      0.1492(17) &      0.6323(18) &    1.275123(11) &      0.03281(2) &   14.5 &     \\
   33 &    0.804553(13) &      0.0720(17) &      0.6807(38) &    1.242926(20) &      0.03220(2) &    7.3 &     \\
   23 &    0.828305(10) &      0.1175(24) &      0.7586(33) &    1.207285(15) &      0.03564(2) &    9.9 &     \\
   18 &    0.854405(08) &      0.1136(17) &      0.3873(23) &    1.170405(11) &      0.03688(2) &   11.7 &     \\
   19 &    0.881783(08) &      0.1139(17) &      0.9447(23) &    1.134066(10) &      0.03634(2) &   11.6 &     \\
    1 &    0.915583(03) &      0.5981(19) &      0.5649(05) &    1.092200(04) &      0.04187(1) &   36.9 &     \\
   10 &    0.954553(06) &      0.1958(17) &      0.0933(14) &    1.047611(07) &      0.04459(1) &   18.3 &     \\
   40 &    0.999928(14) &      0.0556(17) &      0.4315(48) &    1.000072(14) &      0.04754(2) &    6.7 &     \\
   29 &    1.047990(13) &      0.0700(17) &      0.0747(38) &    0.954208(12) &      0.04586(2) &    7.6 & 5$\nu_{5}$-7$\nu_{24}$    \\
   31 &    1.115213(13) &      0.0616(17) &      0.0032(43) &    0.896690(10) &      0.05752(2) &    7.6 & -7$\nu_{14}$+2$\nu_{30}$    \\
    5 &    1.193358(04) &      0.3612(17) &      0.5842(07) &    0.837972(03) &      0.05872(1) &   30.6 &     \\
\hline
\end{tabular}
\end{table*}

(Quasi) regular period spacing pattern predicted by the asymptotic theory enables mode identification and then seismic modelling.
This motivated us to look for such features in the oscillation spectrum of KIC\,3240411. A careful check of period differences
in the range typical for  high-order g modes revealed a period spacing that is shown in Fig.\,\ref{per_spacings_KIC3240411}.
However, there is some ambiguity because there are two possible series that we called, {\it a} and {\it b}. Both series consist
of 22 frequencies and they are listed in Table\,\ref{tab_freq_series}.  The sequence {\it a} can be additionally extended to lower frequencies if we assume that some modes are undetected (grey squares in Fig.\,\ref{per_spacings_KIC3240411}).
As one can see both series include also frequencies that can be classified as combination ones.
But since we are dealing with very dense oscillation spectrum and potential parental frequencies have small amplitudes,
we opt for the statement that fulfilling of mathematical condition on combination frequencies is accidental. Moreover, all frequencies
belonging to the series  have  high $S/N$ ratio. Only for three frequencies this ratio drops slightly below five.
In addition, we identified two shorter series of quasi-equidistant in period frequencies
(marked by cyan and magenta symbols in Fig.\ref{per_spacings_KIC3240411}). However, our seismic modelling
will show that these series are rather accidental. Therefore we do not list their parameters.

The length of the series ({\it a} or {\it b}) as well as the fact that  they involve mostly high-amplitude frequencies
allow us to assume that we observe asymptotic behaviour of g modes. Therefore in our further analysis  we treat  these frequencies 
as modes with the same degree $\ell$, azimuthal order, $m$ and consecutive radial orders $n$. Moreover, based on the slope of the series we can say
that we are dealing with axisymmetric or prograde modes \citep{2017MNRAS.469...13S,2017EPJWC.16003012S}.

\section{Seismic modelling}

\subsection{Standard opacity models}

Seismic modelling was done simultaneously with the mode identification for the series {\it a} and {\it b}.
In that way we verify whether these $\Delta P$ patterns result from asymptotic properties of high-order g modes
or are accidental structures.
We constructed a few extended grids of evolutionary and  non-adiabatic pulsational models and then
compared the theoretical frequencies with the observed ones using  the discriminant $\chi^2$ defined as
\begin{equation}
\chi^2= \frac{1}{n}  \sum_{i=1}^n \frac{(\nu_{\mathrm o,i}-\nu_{\mathrm t,i})^2}{\sigma_i^2},
\end{equation}
where $n$ is the number of consecutive modes, $\nu_{\mathrm o,i}$ are frequencies from the series {\it a} or {\it b},
$\nu_{\mathrm t,i}$ are the theoretical counterparts with a given harmonic degree $\ell$ and azimuthal order $m$.
The observed uncertainties in frequencies, $\sigma_i^2$, were multiplied by a factor 4 \citep[see][]{Moravveji2016}.

Both stellar evolution and pulsation models were calculated within the 3$\,\sigma$ error box in $\log T_\mathrm{eff}$ and $\log g$.
We used MESA code \citep[in its 9575 version,][]{2011ApJS..192....3P, 2013ApJS..208....4P, 2015ApJS..220...15P}  considering
a wide range of the rotational velocity from $V_\mathrm{rot}=40\,\mathrm{km\,s^{-1}}$  on ZAMS
to the upper limit of the validity of the traditional
approximation. As the upper limit we adopted 220 km s$^{-1}$  on ZAMS
that is about a half of the break-up velocity. The step
was $\Delta V_\mathrm{rot}=10\,\mathrm{km\,s^{-1}}$.
We considered various chemical compositions ($X,~Z$), assuming
the element mixture of \citet{2009ARA&A..47..481A} and the OPLIB opacity data \citep{2015HEDP...14...33C, 2016ApJ...817..116C}.
We used MESA tables of the equation of state \citep[see][]{2011ApJS..192....3P} that are based on OPAL EOS \citep{2002ApJ...576.1064R},
nuclear reaction network approx21.net \citep[][]{2011ApJS..192....3P}, and Eddington grey atmosphere  boundary conditions.
More details can be found in the inlist file included in  Appendix D.
The overshooting from the convective core was described by the exponential formula
and its parameter was initially set as $f_\mathrm{ov}=0.02$.
In the first models grid we assumed hydrogen abundance and metallicity as measured in photosphere, i.e.
$X=0.67$, $Z=0.006$.
In all our evolutionary grids we applied a step in mass $\Delta M=0.05\,\mathrm M_{\sun}$  and a maximal step in the effective temperature $\Delta \log T_\mathrm{eff}=0.00066$.

\begin{figure*}
	\includegraphics[width=1.7\columnwidth]{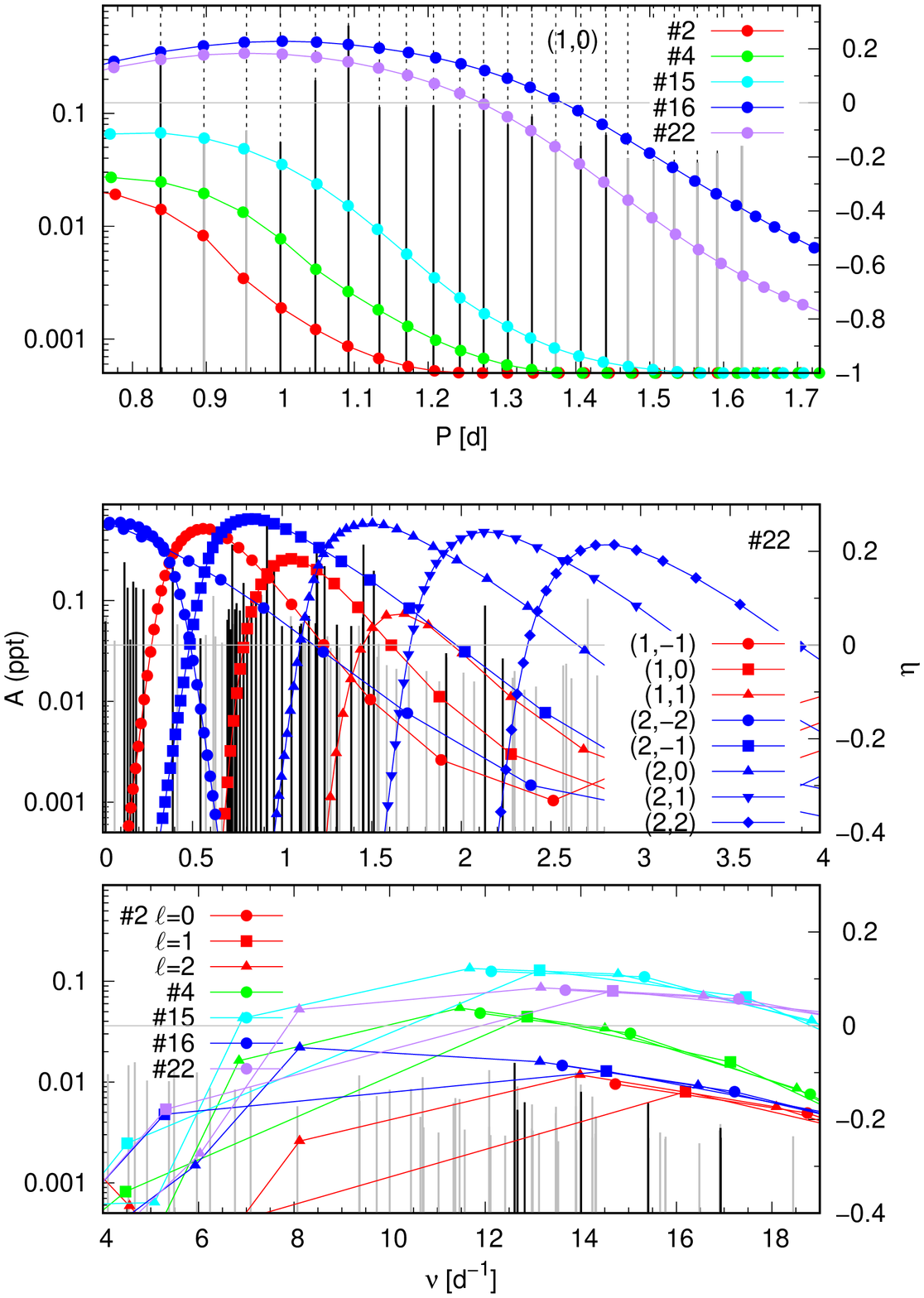}
    \caption{{\it Top panel:} Frequencies and amplitudes for the series {\it b} in comparison with the theoretical frequencies
of axisymmetric dipole modes for selected best seismic models (see Tables\,\ref{tab_best_mod_OPLIB} and \ref{tab_best_mod_OPLIBmod}). The values of the mode instability parameter $\eta$ are on the right Y axis. The observed frequencies are marked
 with black (definitely independent) and grey (possible combination) vertical lines.
  {\it Middle panel:}  The similar plot but all observed frequencies  in the range typical for g-mode pulsations were compared with dipole and quadrupole modes of  the model \#22.
 {\it Bottom panel:} The observed frequencies and their amplitudes in the  p-mode range confronted with the theoretical frequencies
 of models considered in the top panel.}
    \label{best_models}
\end{figure*}

\begin{figure*}
	\includegraphics[width=1.235\columnwidth, angle=270]{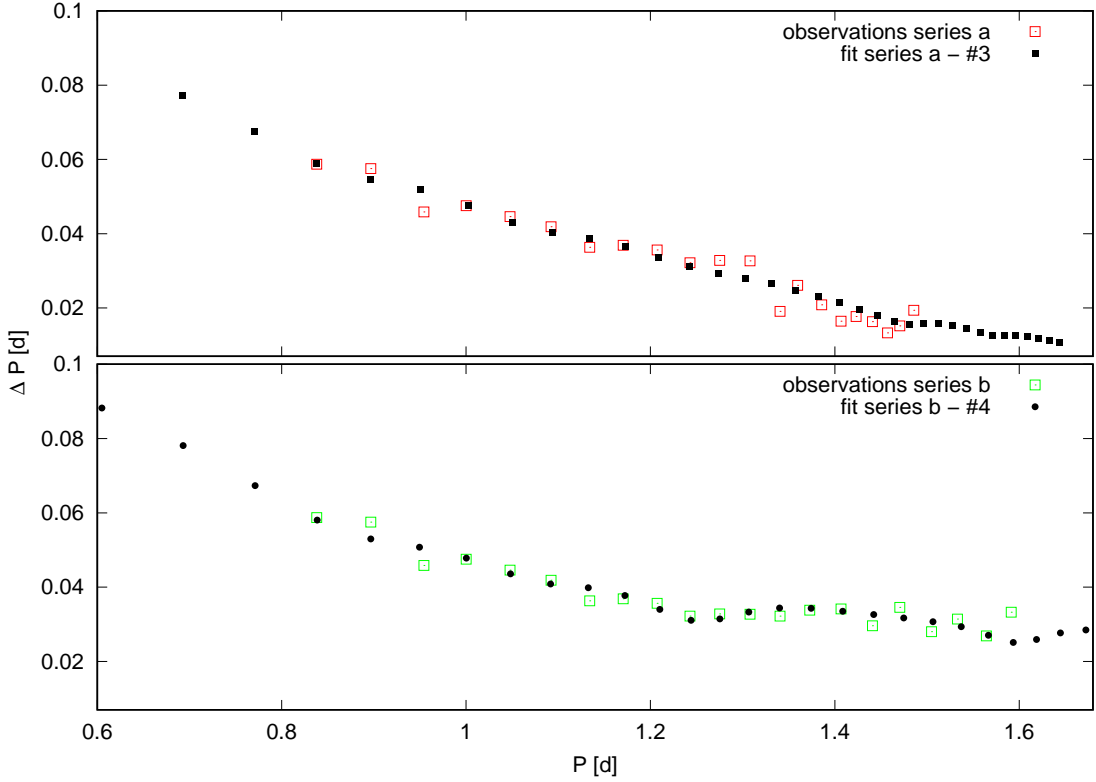}
    \caption{A comparison of the quality of a fit for models \#3 and \#4 (see the text for details). The values of the observed and theoretical
               period spacings are shown as a function of period.}
    \label{best_models_solar}
\end{figure*}

\begin{figure*}
	\includegraphics[width=1.235\columnwidth, angle=270]{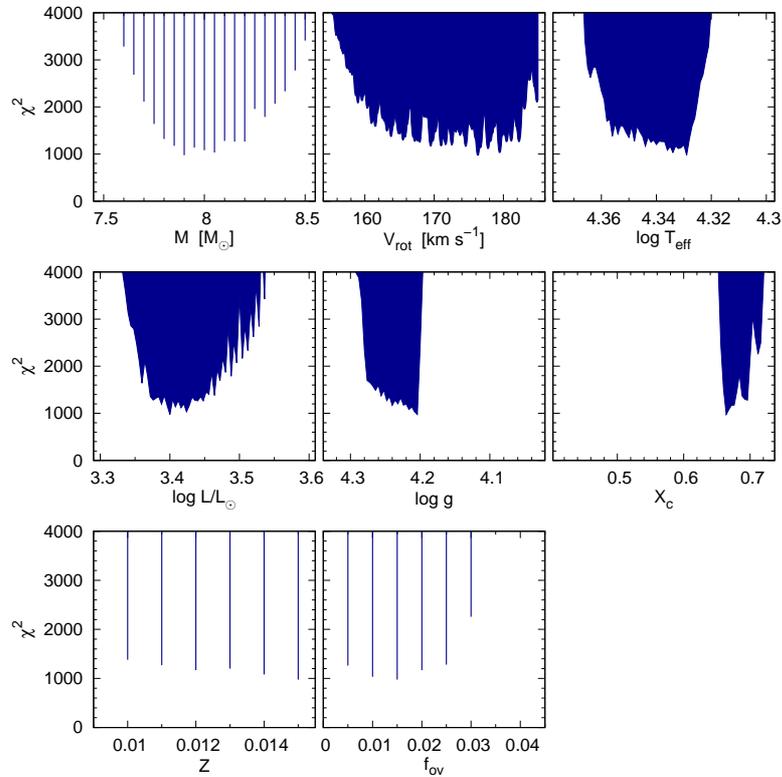}
    \caption{The values of $\chi^2$ as a function of various parameters in the vicinity of the best model \#4 (see text for details).}
    \label{chi_OPLIB_set1}
\end{figure*}

\begin{figure*}
	\includegraphics[width=1.235\columnwidth, angle=270]{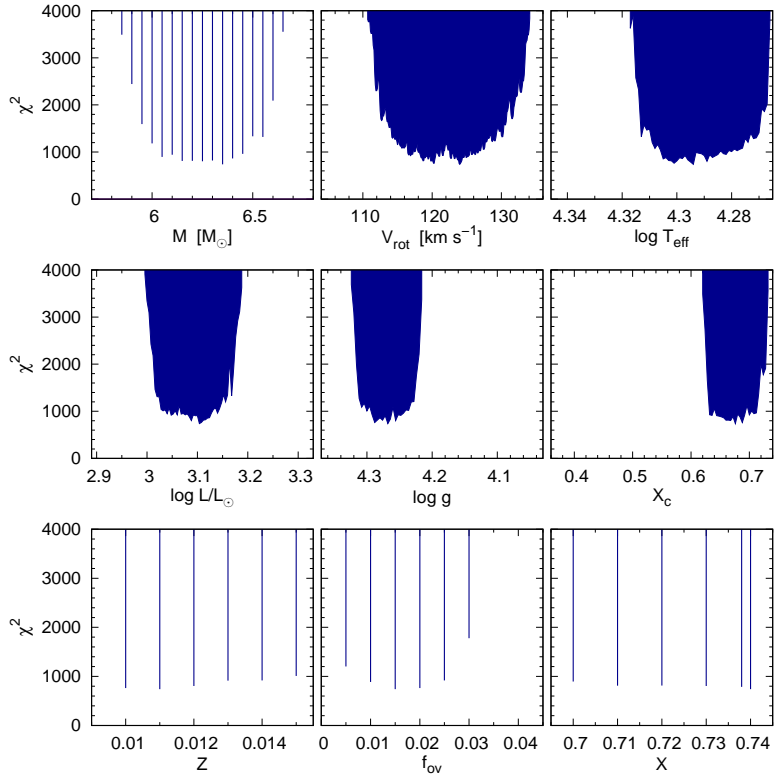}
    \caption{The same as in Fig.\ref{chi_OPLIB_set1} but in the vicinity of the model \#19  and the dependence on the initial hydrogen abundance, $X$, was added.}
    \label{chi_OPLIB_set6}
\end{figure*}

\begin{table*}
	\centering
	\caption{The parameters of the best seismic models reproducing the series {\it a} and {\it b} of the period spacing.
	Because of very high values of $\chi^2$ for $\ell=3$ only a few examples were included.
	The subsequent columns contain: the name of the series, mass, the current value of rotation, the rotation frequency, the initial abundance of hydrogen, metallicity, overshooting parameter,
	effective temperature, luminosity, surface gravity, central hydrogen abundance, angular numbers, $\chi^2$ discriminant,
	the maximum value of the $\eta$ parameter for the modes associated with the observed series. In the penultimate column,
	we give the number of unstable theoretical modes associated with the observed series. In the last column, selected models have been marked with identifiers.}
	\label{tab_best_mod_OPLIB}
	\setlength{\tabcolsep}{5.7pt}
	\begin{tabular}{cccccccccccccccc}
\hline	
S. & M     & $V_\mathrm{rot}$ & $\nu_\mathrm{rot}$ & $X$ & $Z$ & $f_\mathrm{ov}$ &  $\log T_\mathrm{eff}$ & $\log L/\mathrm L_{\sun}$ & $\log g$ & $X_\mathrm c$ & ($\ell$, $m$) & $\chi^2$ & $\eta_\mathrm{max}$ & N & mod. \\
 & [M$_{\sun}$]     & $[\mathrm{km}\,\mathrm{s}^{-1}$] &  [d$^{-1}$] & &  &  &   &    &  &  &  &  &  &  &  \\
	\hline

a & 7.60 & 223.35 & 0.91 & 0.670 & 0.006 & 0.02  & 4.35230 & 3.7346 & 3.947 & 0.289 &   (1,   0)  &        41555 & -0.24 &  0  &         \\
a & 7.95 & 136.30 & 0.42 & 0.670 & 0.006 & 0.02  & 4.33116 & 3.9011 & 3.715 & 0.109  &  (1,  1)  &         7282 & -0.34 &  0  &  \#1       \\

b & 6.25 & 117.35 & 0.82 & 0.670 & 0.006 & 0.02  & 4.34864 & 3.2550 & 4.327 & 0.612  &  (1, 0)  &               1534 & -0.39 &  0  &     \#2    \\
b & 4.85 & 40.90  & 0.34 & 0.670 & 0.006 & 0.02  & 4.29161 & 2.8792 & 4.364 & 0.623 &   (1,   1)  &                  7084 & -0.48 &  0  &         \\

\hline

a & 6.65 & 159.20 & 0.87 & 0.738 & 0.013 & 0.02  & 4.28820 & 3.2230 & 4.144 & 0.577 &   (1,   0)        &  49442 & -0.09 &  0  &         \\
a & 8.55 & 82.55  & 0.47 & 0.738 & 0.013 & 0.02  & 4.36613 & 3.5038 & 4.284 & 0.716 &   (1,   1)   &        3782 & -0.53 &  0  &    \#3     \\

b & 7.95 & 171.65 & 0.95 & 0.738 & 0.013 & 0.02  & 4.34017 & 3.4212 & 4.231 & 0.676  &  (1,  0)    &     1201 & -0.29 &  0  &      \#4 \\
b & 6.20 & 52.85  & 0.38 & 0.738 & 0.013 & 0.02  & 4.29661 & 3.0208 & 4.349 & 0.737  &  (1,  1)         &  18340 & -0.42 &  0  &         \\

\hline

a & 6.55 & 195.60 & 0.87 & 0.670 & 0.013 & 0.02 & 4.29170 & 3.4146 & 3.960 & 0.367   & (1, 0)  & 49306 & -0.02  & 0    &         \\
b & 7.40 & 165.75 & 0.94 & 0.670 & 0.013 & 0.02 & 4.35001 & 3.4384 & 4.222 & 0.605 &   (1,   0)  & 1212 & -0.32 &  0 &     \#5 \\
\hline
a & 6.50 & 157.85 & 0.87 & 0.710 & 0.013 & 0.02 & 4.29408 & 3.2370 & 4.144 & 0.554  &  (1,  0)  & 48601 & -0.11 &  0 &         \\
b & 7.75 & 169.35 & 0.95 & 0.710 & 0.013 & 0.02 & 4.34520 & 3.4313 & 4.230 & 0.649  &  (1,   0)  & 1224 & -0.31 &  0 &    \#6 \\

\hline
a & 7.25 & 74.40  & 0.44 & 0.670 & 0.006 & 0.02 & 4.37560 & 3.5072 & 4.241 & 0.526 &   (2,   0)   &       35235 & -0.71  &  0    &         \\
a & 9.85 &  72.95 & 0.19 & 0.670 & 0.006 & 0.02 & 4.37042 & 4.2134 & 3.653 & 0.076  &  (2,   1)   &      172049 & -0.29 &  0  &         \\
a & 10.10 & 46.45 & 0.12 & 0.670 & 0.006 & 0.02 & 4.37605 & 4.2465 & 3.653 & 0.077  &  (2,   2)    &     195992 & -0.32 &  0   &         \\
b & 9.25 & 126.80 & 0.35 & 0.670 & 0.006 & 0.02 & 4.35905 & 4.0993 & 3.690 & 0.101 &   (2,  0)    &   15806 & -0.18 &  0 &         \\
b & 8.55 & 107.70 & 0.34 & 0.670 & 0.006 & 0.02 & 4.36035 & 3.9783 & 3.786 & 0.157   & (2,  1) &  43956 & -0.24 &  0 &         \\
b & 9.80 &  40.55 & 0.10 & 0.670 & 0.006 & 0.02 & 4.37012 & 4.2079 & 3.655 & 0.075  &  (2,  2)  &    70247 & -0.30  & 0 &         \\

\hline
a & 10.85 & 176.80 & 0.45 & 0.738 & 0.013 & 0.02 &  4.34768 & 4.1290 & 3.688 & 0.221  &  (2,   0)   &       23135 & 0.02 &  6            &         \\

a & 12.50 & 109.50 & 0.25 & 0.738 & 0.013 & 0.02 &  4.37753 & 4.3306 & 3.668 & 0.205  &  (2,   1)   &       27596 & -0.23 & 0            &         \\

a & 8.65 &  38.9   & 0.16 & 0.738 & 0.013 & 0.02 & 4.34699 & 3.6813 & 4.035 & 0.466  &  (2,  2)     &     72963 & -0.40  & 0   &         \\

b & 9.00 &  78.05  & 0.41 & 0.738 & 0.013 & 0.02 & 4.37385 & 3.5979 & 4.243 & 0.683 &   (2,  0)  &           6240 & -0.67 &  0   &         \\

b & 12.15 & 101.50 & 0.23 & 0.738 & 0.013 & 0.02 & 4.37036 & 4.2976 & 3.660 & 0.196 &   (2,   1)  &         9832 & -0.19 &  0   &         \\

b & 12.55 &  66.40 & 0.15 & 0.738 & 0.013 & 0.02 &  4.37748 & 4.3422 & 3.658 & 0.195 &   (2,  2)    &  15129 & -0.28 &  0  &         \\

\hline

a & 7.00 &  45.25 & 0.30 & 0.670 & 0.006 & 0.02 & 4.37701 & 3.4126 & 4.332 & 0.620 &   (3, 0)    &     253259 & -1.00 &  0  &         \\

b & 7.00 &  45.25 & 0.30 & 0.670 & 0.006 & 0.02 & 4.37701 & 3.4126 & 4.332 & 0.620   & (3, 0)       &   64842 & -1.00 &  0 &         \\

\hline

a & 6.55 &  44.10 & 0.26 & 0.738 & 0.013 & 0.02 & 4.29411 & 3.1671 & 4.217 & 0.636 &   (3, 0)    &  91965 & -0.39 &  0   &         \\

b & 12.45 &128.95 & 0.30 & 0.738 & 0.013 & 0.02 & 4.37658 & 4.3224 & 3.670 & 0.209  &  (3, 0) &  23280 & -0.02 &  0 &         \\

\hline
b & 7.90 & 176.30 & 0.95 & 0.738 & 0.015 & 0.015 & 4.32947 & 3.4014 & 4.205 & 0.665  &  (1,    0) &  975 & -0.24 &  0  & \#7\\

\hline
\end{tabular}
\end{table*}

\begin{table*}
	\centering
	\caption{The opacity modification coding.}
	\label{tab_opac_modifications}
	\begin{tabular}{ccccccccccccc}
\hline	
\# &  $\log T_{0,1}$ & $a_1$ & $b_1$ & $\log T_{0,2}$ & $a_2$ &  $b_2$ & $\log T_{0,3}$ & $a_3$  & $b_3$ & $\log T_{0,4}$ & $a_4$ & $b_4$ \\
	\hline
O1  & 5.30 & 0.1 & 1.0  &      &     &     &      &     &     &      &     &   \\
O2  & 5.46 & 0.1 & 1.0  &      &     &     &      &     &     &      &     &   \\
O3  & 5.30 & 0.1 & 1.0  & 5.46 & 0.1 & 1.0 &      &     &     &      &     &  \\
O4  & 5.06 & 0.1 & 1.0  & 5.30 & 0.1 & 1.0 & 5.46 & 0.1 & 1.0  &      &     &  \\
O5  & 5.46 & 0.1 & 2.0  &      &     &     &      &     &     &      &     &   \\
O6  & 5.06 & 0.1 & 0.5  & 5.22 & 0.1 & 0.5 & 5.30 & 0.1 & 1.0 & 5.46 & 0.1 & 2.0 \\
O7  & 5.06 & 0.1 & -0.5 & 5.22 & 0.1 & 0.5 & 5.30 & 0.1 & 1.0 & 5.46 & 0.1 & 2.0\\
O8  & 5.06 & 0.1 & -0.5 & 5.22 & 0.1 & 0.5 & 5.46 & 0.1 & 2.0 &      &     &  \\
O9  & 5.06 & 0.1 & 0.5  & 5.22 & 0.1 & 0.5 & 5.46 & 0.1 & 2.0 &      &     &  \\
O10 & 5.22 & 0.1 & 0.5  & 5.46 & 0.1 & 2.0 &      &     &     &      &     & \\
\hline
\end{tabular}
\end{table*}

\begin{table*}
	\centering
	\caption{Same as in Table\,\ref{tab_best_mod_OPLIB} but for grids with the modified opacities and only for dipole modes.}
	\label{tab_best_mod_OPLIBmod}
	\setlength{\tabcolsep}{4.9pt}
	\begin{tabular}{ccccccccccccccccc}
\hline	
S. & M     & $V_\mathrm{rot}$ & $\nu_\mathrm{rot}$ & $X$ & $Z$ & $f_\mathrm{ov}$ &  $\log T_\mathrm{eff}$ & $\log L/\mathrm L_{\sun}$ & $\log g$ & $X_\mathrm c$ & $(\ell,\,m)$ & $\chi^2$ & $\eta_\mathrm{max}$ & N & opac & mod. \\
 & [M$_{\sun}$]     & $[\mathrm{km}\,\mathrm{s}^{-1}$] & [d$^{-1}$] &  &  &  &   &    &  &  &  &  &  &  &  &\\
	\hline

a & 7.60 & 224.55 & 0.91 & 0.670 & 0.006 & 0.02 &  4.35087 & 3.7360 & 3.940 & 0.287 &   (1,  0)  &        41773 & -0.41 &  0     &  O1 & \\
a & 7.00 &  64.85 & 0.44 & 0.670 & 0.006 & 0.02 & 4.37763 & 3.4018 & 4.345 & 0.635  &  (1,   1)       &    6442 & -0.91 &  0    &  O1 & \\
b & 5.20 &  92.40 & 0.72 & 0.670 & 0.006 & 0.02  & 4.30630 & 2.9855 & 4.347 & 0.617 &   (1,  0)  &      2153 & -0.65 &  0 &  O1 & \#8 \\
b & 4.85 &  41.00 & 0.34 & 0.670 & 0.006 & 0.02 &  4.29114 & 2.8796 & 4.362 & 0.623 &   (1,  1)  & 7388 & -0.81 &  0 &  O1 & \\
\hline
a & 6.65 & 159.35 & 0.87 & 0.738 & 0.013 & 0.02 &  4.28766 & 3.2230 & 4.142 & 0.578  &  (1,   0)   &       44117 & -0.16 &  0   &       O1 &\\
a & 8.55 &  82.70 & 0.47 & 0.738 & 0.013 & 0.02 &  4.36568 & 3.5036 & 4.282 & 0.716  &  (1,   1)   &        3629 & -0.74 &  0   &       O1 &\\
b & 7.90 & 170.20 & 0.94 & 0.738 & 0.013 & 0.02   & 4.33816 & 3.4119 & 4.230 & 0.676 &   (1,  0)  &         1784 & -0.48 &  0 &       O1 & \#9\\
b & 6.10 &  51.80 & 0.37 & 0.738 & 0.013 & 0.02 & 4.29240 & 2.9968 & 4.350 & 0.737  &  (1,  1)  & 18871 & -0.65 &  0 &       O1 \\

\hline

a & 7.55 & 222.05 & 0.90 & 0.670 & 0.006 & 0.02 & 4.34916 & 3.7264 & 3.940 & 0.287 &   (1,  0)       &   34618 & 0.14 & 14    &      O2 & \\
a & 7.95 & 137.40 & 0.42 & 0.670 & 0.006 & 0.02 & 4.32963 & 3.9003 & 3.710 & 0.111  &  (1,  1)       &    6692 & -0.01 &  0  &      O2 & \\
b & 6.15 & 112.60 & 0.79 & 0.670 & 0.006 & 0.02 &  4.34485 & 3.2308 & 4.329 & 0.614  &  (1, 0) & 1091 & -0.09 &  0  &      O2 & \#10\\
b & 5.00 &  46.00 & 0.34 & 0.670 & 0.006 & 0.02 & 4.29253 & 2.9787 & 4.282 & 0.544  &  (1,  1)  &    7141 &  0.04 &  2 &      O2  &\\
\hline
a & 6.60 & 156.85 & 0.86 & 0.738 & 0.013 & 0.02 & 4.28565 & 3.2103 & 4.143 & 0.580 &   (1,   0)     &     46283 &  0.20 & 15 &      O2 & \\
a & 8.50 &  82.35 & 0.46 & 0.738 & 0.013 & 0.02 & 4.36362 & 3.4970 & 4.278 & 0.713 &   (1,  1)   & 3103 & -0.49 &  0 &      O2  &\\
b & 7.85 & 165.85 & 0.93 & 0.738 & 0.013 & 0.02  & 4.33691 & 3.3993 & 4.235 & 0.681 &   (1,   0)          & 1054 & 0.02 &  3 &      O2  & \#11\\
b & 5.80 &  55.15 & 0.38 & 0.738 & 0.013 & 0.02 & 4.26719 & 2.9469 & 4.277 & 0.691  &  (1,  1)  &  14591 & 0.11 &  4 &      O2 & \\
\hline

a & 7.55 & 222.90 & 0.90 & 0.670 & 0.006 & 0.02 & 4.34794 & 3.7273 & 3.934 & 0.286 &   (1,   0)      &    32885 & 0.07 &  8 &     O3 & \\
a & 7.95 & 138.20 & 0.42 & 0.670 & 0.006 & 0.02 & 4.32825 & 3.9005 & 3.704 & 0.110  &  (1,  1)   &  6469 & -0.06 &  0 &     O3 & \\
b & 6.15 & 112.70 & 0.79 & 0.670 & 0.006 & 0.02 & 4.34445 & 3.2307 & 4.327 & 0.614  &  (1,  0)  &   1200 & -0.26  & 0 &     O3  & \#12\\
b & 5.00 &  46.15 & 0.34 & 0.670 & 0.006 & 0.02 & 4.29167 & 2.9796 & 4.277 & 0.542  &  (1,  1)       &   6916 & -0.14 &  0 &     O3 &\\

\hline
a & 6.90 & 198.65 & 0.86 & 0.738 & 0.013 & 0.02 & 4.27495 & 3.3687 & 3.961 & 0.423   & (1,  0)     &     43225 & 0.27 & 20 &     O3 &\\
a & 8.50 &  82.70 & 0.46 & 0.738 & 0.013 & 0.02 & 4.36299 & 3.4976 & 4.275 & 0.712 & (1, 1)   &    2926 & -0.65 &  0 &     O3 \\
b & 7.80 & 164.50 & 0.92 & 0.738 & 0.013 & 0.02  & 4.33532 & 3.3917 & 4.233 & 0.680 &   (1,   0)          &   1214 & -0.06 &  0 &     O3 & \#13\\
b & 5.80 &  55.20 & 0.38 & 0.738 & 0.013 & 0.02 & 4.26677 & 2.9471 & 4.275 & 0.691 &  (1, 1) &   13464 & 0.03 &  2 &     O3 &\\

\hline

a & 7.55 & 223.45 & 0.90 & 0.670 & 0.006 & 0.02 & 4.34762 & 3.7258 & 3.934 & 0.288 & (1, 0) & 32636 & 0.03 &  6 &     O4 &\\
a & 8.00 & 140.80 & 0.42 & 0.670 & 0.006 & 0.02 & 4.32743 & 3.9111 & 3.693 & 0.107 & (1, 1)  & 6927 & -0.09 &  0 &     O4 &\\
b & 6.15 & 113.25 & 0.79 & 0.670 & 0.006 & 0.02 & 4.34398 & 3.2321 & 4.324 & 0.612 &  (1, 0)  &  1172 & -0.29 &  0 &     O4 & \#14\\
b & 5.00 &  46.15 & 0.34 & 0.670 & 0.006 & 0.02 & 4.29167 & 2.9796 & 4.277 & 0.542 & (1, 1) &  6916 & -0.14 &  0 &     O4 &\\

\hline
a & 6.90 & 199.65 & 0.86 & 0.738 & 0.013 & 0.02 & 4.27399 & 3.3694 & 3.957 & 0.422  &  (1,  0)    &      41642 & 0.22 & 20 &     O4 &\\
a & 8.35 &  81.70 & 0.46 & 0.738 & 0.013 & 0.02 & 4.35812 & 3.4732 & 4.272 & 0.711 & (1, 1)  & 2887 & -0.66 &  0 &     O4 &\\
b & 7.80 & 164.35 & 0.92 & 0.738 & 0.013 & 0.02  & 4.33503 & 3.3907 & 4.233 & 0.681 &   (1,   0)          &   1133 & -0.11 &  0 &     O4 & \#15\\
b & 5.80 &  55.20 & 0.38 & 0.738 & 0.013 & 0.02 & 4.26644 & 2.9468 & 4.274 & 0.691 & (1, 1)  & 13334 & -0.02 &  0 &     O4 &\\

\hline
b & 6.25 & 124.05 & 0.80 & 0.738 & 0.013 & 0.02 & 4.28317 & 3.0616 & 4.258 & 0.686 &   (1,  0)  &    1454 & 0.23 & 14 & O5 &  \#16\\
\hline
b & 6.20 & 122.35 & 0.79 & 0.738 & 0.013 & 0.02 & 4.28022 & 3.0497 & 4.255 & 0.685  &  (1,   0)  &     984 & 0.15 &  9 & O6 &  \#17\\
\hline
b & 6.25 & 123.75 & 0.80 & 0.738 & 0.013 & 0.02 & 4.28303 & 3.0614 & 4.258 & 0.686  &  (1,   0)  &    1504 & 0.22 & 12 & O7 &  \#18\\
\hline
b & 6.25 & 123.85 & 0.80 & 0.738 & 0.013 & 0.02 & 4.28320 & 3.0615 & 4.259 & 0.686  &  (1,  0)      & 1344 & 0.21 & 12& O8 & \#19\\
\hline
b & 6.25 & 124.30 & 0.80 & 0.738 & 0.013 & 0.02 & 4.28259 & 3.0620 & 4.256 & 0.685  &  (1, 0)   &  1420 & 0.18 & 11 & O9 & \#20\\
\hline
b & 6.25 & 124.00 & 0.80 & 0.738 & 0.013 & 0.02 & 4.28286 & 3.0615 & 4.257 & 0.686   & (1,   0) &  1279 &  0.19 & 11 & O10 & \#21\\
\hline
b & 6.35 & 124.00 & 0.80 & 0.740 & 0.011 & 0.015 & 4.29416 & 3.1046 & 4.266 & 0.676  &  (1,    0) &  740 & 0.18  & 10 & O8 &\#22\\
\hline

\end{tabular}
\end{table*}

Because the frequencies of g modes are often of the order of the rotation frequency, including the effects of rotation on pulsations is vital.
This has been done in the framework of the traditional approximation \citep{1996ApJ...460..827B,1997ApJ...491..839L,2003MNRAS.340.1020T, 2005MNRAS.360..465T}
which takes into account the effects of the Coriolis force. The values of the rotation frequency for our best seismic models
are listed in the fourth column of Table\,\ref{tab_best_mod_OPLIB}  and \ref{tab_best_mod_OPLIBmod}.
All axisymmetric and prograde modes with $\ell \le3$ were examined. This part of calculations was made with the use
of the Dziembowski code in its version that employs the traditional approximation \citep{2007MNRAS.374..248D}.
In pulsational computations we decreased the step in $V_\mathrm{rot}$ to 0.05 km s$^{-1}$ and the lower limit of $V_\mathrm{rot}$
was set as 35 km s$^{-1}$. This is below the minimum surface velocity resulting from $V_\mathrm{rot}\sin i$ but we wanted to check if the core
can rotate slower than the surface layers.

Based on the values of $\chi^2$ we can safely reject the hypothesis that considered series are $\ell=2$ or 3 modes (see Table \,\ref{tab_best_mod_OPLIB}).
Our fitting points that observations are best reproduced under the assumption that the series {\it b}
consists of frequencies with the consecutive radial orders $n$  associated with the ($\ell=1$, $m=0$) modes.
For the series {\it a} we got the best fit for prograde dipole modes ($\ell=1,~m=+1$) but the values $\chi^2$ for model \#1
is more than three times larger than for the best fit of the series $b$  (model \#2).

As usually, we estimate the mode instability using the dimensionless, normalized parameter $\eta$ defined by the work integral $W$ over the pulsational cycle \citep{1978AJ.....83.1184S},
\begin{equation}
\eta= \frac{W}{\int_0^R \left|\frac{dW}{dr}\right| dr}.
\end{equation}
The values of $\eta$ range from -1 to 1 and  unstable modes have $\eta>0$. Unfortunately, all modes from both series are far from being unstable.
In the best model (\#2 in Table\,\ref{tab_best_mod_OPLIB}) the maximum value of $\eta$ for frequencies that reproduce observations reaches $-0.39$ and many modes are entirely stable with $\eta\approx-1$. In the top panel of Fig.\,\ref{best_models} the values of $\eta$
for this model are plotted as  red dots.

Since the metallicity is crucial for pulsation excitation in B-type stars one can expect that increasing $Z$ improves instability conditions.
We calculated the second grid of models with the solar abundances \citep{2009ARA&A..47..481A}, i.e., $X=0.738$ and $Z=0.013$.
As before, the series $a$ prefers the $\ell=1,~m=+1$  modes and the series $b$  - the $\ell=1,~m=0$ modes.
The values of $\chi^2$ got smaller for both series but again the solutions for the series $b$ are strongly favored.
In Fig.\,\ref{best_models_solar} we compare the best seismic models for the series $a$ and $b$ with observations on the diagram $\Delta P$ vs. $P$.
The parameters of the seismic models for the series $a$ and $b$ are listed in the sixth and seventh line of Table\,\ref{tab_best_mod_OPLIB} (models \#3 and \#4), respectively.
For the series $b$ the parameter $\eta$ increased but modes are still stable
($\eta_\mathrm{max}= -0.29$). In the case of  the series {\it a} , it slightly decreased ($\eta_\mathrm{max}=-0.53$) which is a consequence 
of the shift of the maximum of $\eta(\nu)$ towards higher frequencies.
The values of $\eta$ for the best seismic model from this grid (the model \#4) are plotted in Fig.\,\ref{best_models} as green dots.

We also checked two more grids with a solar metallicity: for the hydrogen abundances as determined by \citet{2012A&A...539A.143N} for B-type stars
in the neighborhood of the Sun and for helium abundance as determined for our target by \citet{Lehmann2011}.
Also these calculations confirmed our previous conclusions. The values of $\chi^2$
are comparable to the previous results and  choose the series $b$  with axisymmetric dipole modes.
However, changing the grids did not solve the problem with mode excitation (see Table\,\ref{tab_best_mod_OPLIB}).
On the other hand, this should not be surprising because the star is outside or on the edge of the SPB instability strip calculated for standard opacity data and reasonable metallicity \citep[see][]{2017MNRAS.469...13S}.

In the next step, we tried to constrain the metallicity and overshooting parameter.
To this end we fitted the (1,0) modes to the series $b$ for parameters in the vicinity of the model \#4, i.e. we constructed a fine grid of
models with $M$ from 7.5 to 8.5\,$\mathrm M_{\sun}$, $V_\mathrm{rot}$ from 155 to 185 km s$^{-1}$, X=0.738, $Z$ from 0.010 to 0.015 with $\Delta Z=0.001$ and $f_\mathrm{ov}$ from 0.005 to 0.040 with $\Delta f_\mathrm{ov}=0.005$. The parameters of the best model (\#7)
from this grid are listed in the last line of Table\,\ref{tab_best_mod_OPLIB}
and the values of $\chi^2$ as a function of different parameters are shown in Fig.\,\ref{chi_OPLIB_set1}. As one can see the overshooting parameter should not be higher than 0.03. Moreover, the star is at the beginning of its course on the main sequence evolution.
This conclusion can be drawn also from the four earlier grids.
Because the model \#7 from the fine grid has $Z=0.015$ it has slightly better instability conditions.
All best seismic models that fit the series $b$ were marked on the Kiel diagram in Fig.\,\ref{Kiel} with big symbols. As one can see
all of them are within the $3\sigma$ error box.

The two other series shown in Fig.\,\ref{per_spacings_KIC3240411}  with $\Delta P\approx 0.01$  cannot belong
to  any asymptotic pattern and are composed of modes with different angular numbers  $(\ell,\,m)$.

\subsection{Modified opacity models}

The problem with excitation of g modes in hybrid early B-type pulsators is known for many years \citep{2004MNRAS.350.1022P}.
Recent results for the early B-type star $\nu$ Eridani by \citet{2017MNRAS.466.2284D} showed that significant modifications 
 of the mean opacity profile are necessary
to account for all pulsational properties. In particular, to excite high-order g modes they had to increase the mean opacity
at $\log T=5.46$ by more than 150\%.
This solution appeared to be quite common and works also for other hybrid B-type pulsators \citep{2017sbcs.conf..138D}.
At the depth $\log T=5.46$ nickel has its maximum contribution to opacities. The other modification considered are:
$\log T=5.3$ -- the Z-bump, $\log T=5.22$ -- the maximum contribution of chromium and manganese and $\log T=5.06$ -- the new opacity bump
identified in Kurucz models of stellar atmospheres by \citet{2012A&A...547A..42C}.

Here we follow this line and modify the standard OPLIB opacities as was done by  \citet{2017MNRAS.466.2284D}:
\begin{equation}
\kappa(T)=\kappa_0(T)\left[1+ \sum_{i=1}^{N} b_i \times \exp \left( - \frac{\left( \log T- \log T_{0,i} \right)^2}{a_i^2} \right)  \right]
\end{equation}
where $\kappa_0(T)$ is the standard mean opacity profile and $(a,\, b,\, T_0)$ are parameters
of a Gaussian describing the width, height and position of the maximum, respectively.
Our modifications are coded as OX$(T_{0, i},\,a_i,\, b_i)$ and are listed in Table\,\ref{tab_opac_modifications}.

Our goal was to explain instability of both  g and p modes. The results are listed in Table\,\ref{tab_best_mod_OPLIBmod}.
We stress that throughout the  paper the effects of rotation on pulsation are included only for g modes
and for p modes they are ignored. Both series, $a$ and $b$, were considered.
For O1--O4 modifications we made calculation for two sets of abundances -- the observed one and the solar one.
In these cases, we increased opacities by 100\% at one to three depths expressed by $\log T$.
 All calculations confirmed that the series $b$ is associated with (1,0) modes
and the series $a$,  if considered as a real one, with (1, $+1$) modes. However, O1--O4 modifications do not allow us to excite appropriate modes.
In the best case, which is the O2 modification, we obtained only three unstable modes associated with the observed frequencies from the series $b$.

In the next set of modifications, O5--O10, we substantially increased opacities at $\log T=5.46$ which
allowed to excite many g modes. In the case of O5 we found 14 unstable modes (the model \#16).
Unfortunately in this case all p modes are stable (see the top and bottom panel of Fig.\,\ref{best_models}, respectively).
The minor modification at $\log T=5.30,\,5.22$, and 5.06 allowed us to find models with unstable modes both in g- and p-mode regime (models \#17, \#18, \#19, and marginally \#21).

In the vicinity of the model \#19 we calculated a fine grid of models.
This model was chosen because it has relatively small $\chi^2$ and large number of unstable modes reproducing the series $b$
as well as unstable p-modes. This fine grid contains models calculated
with O8 modifications and $M$ from 5.75 to 6.75\,$\mathrm M_{\sun}$, $V_\mathrm{rot}$ from 105 to 135 km s$^{-1}$, X from 0.70 to 0.74 with $\Delta X=0.1$ (with one additional value of X=0.738),
$Z$ from 0.010 to 0.015 with $\Delta Z=0.001$, and $f_\mathrm{ov}$ from 0.005 to 0.040 with $\Delta f_\mathrm{ov}=0.005$.
The parameters of the best model (\#22) from this grid are listed in Table\,\ref{tab_best_mod_OPLIBmod}.
The instability parameter $\eta$ of the model \#22 is depicted in Fig.\,\ref{best_models}.

The values of $\chi^2$ as a function of different parameters in this fine grid of models are shown in Fig.\,\ref{chi_OPLIB_set6}.
Again we can conclude that in order to reproduce the observations we have to exclude high values of the overshooting parameter,
we got $f_\mathrm{ov}<0.03$, and that the star is near ZAMS.

The Correlation between various parameters of seismic models are discussed in Appendix C.

\section{Conclusions}

We performed seismic modelling of the early B-type star KIC\,3240411 that is a hybrid pulsator
whose oscillation spectrum is dominated by g-mode frequencies.
This modelling is based on the (quasi) regular period spacing identified in the low-frequency range.
Till now, this is the hottest SPB star for which we observe the asymptotic period spacing for g modes.

Applying the Fourier frequency analysis we extracted 389 frequency peaks with $S/N\ge 4$. The more rigorous condition $S/N\ge 5$
gave 72 frequency peaks. Using various customized masks we show that all extracted frequencies are associated with the target star.
The frequencies and their amplitudes vary  in a short time scale which can result from mode interactions and nonlinear effects.

In the low-frequency range of the oscillation spectrum of KIC\,3240411 we identified two series consisting of 22 frequencies
that form the period spacing structures.
We computed extended grids of models using MESA code for evolutionary calculations and linear non-adiabatic code for pulsational
calculations. The effects of rotation on g-mode pulsations were included via the traditional approximation.
Seismic modelling made simultaneously with the mode identification proved the asymptotic character of one of the two identified series.
The best solution was obtained with the series corresponding to the consecutive axisymmetric dipole modes.
This result differs from the previous ones. In the case of other  rotating stars with period spacing, these structures were usually
associated with prograde dipole modes.

We also found constraints on the overshooting parameter, $f_\mathrm{ov} \le 0.03$., and the central value of the hydrogen abundance, $X_c>0.6$.
 It means that the star is rather close to ZAMS, i.e. at the beginning of its main-sequence evolution.
This conclusion is independent of the adopted opacity data.

In the next step, we investigated the mode instability in our seismic models.
All models constructed with the standard opacity tables have stable pulsational modes in the low-frequency range.
The excitation problem occurs also in the  higher frequency range corresponding to p modes.
To solve this problem we followed the approach of \citet{2017MNRAS.466.2284D} and modified the standard mean opacities.
We considered modification at the depths, expressed in $\log T$, at which iron-group elements have their main contributions.

Seismic modelling of KIC\,3240411 clearly shows that there is still a room to improve the opacity data.
Our analysis confirmed that a large increase of opacity at $\log T=5.46$ is necessary to excite high-order g modes.
To get unstable p modes in the observed frequency range, we had to enhance the mean opacities
at $\log T=5.3$ or $\log 5.22$.  In some cases, the opacities had to be extra decreased at $\log T=5.06$
to excite the highest frequency modes.

At the moment there is no good explanation for these opacity modifications. Their need may result from some inaccuracies in the opacity
calculation methods or can be intrinsically related to a star, e.g., non-homogenous distribution of chemical elements in the interior.
This issue requires further in-depth studies and seismic modelling of hybrid pulsators  seems to be one
of the most stringent test of these microphysics data.

\section*{Acknowledgements}
 This work was financially supported by the Polish National Science Centre grant
2015/17/B/ST9/02082. Calculations have been carried out
using resources provided by Wroclaw Centre for Networking
and Supercomputing (http://wcss.pl), grant no. 265.
Funding for the
Kepler
mission is provided by the NASA Science Mission directorate. Some of
the data presented in this paper were obtained from the Multimission Archive
at the Space Telescope Science Institute (MAST). STScI is operated by the
Association of Universities for Research in Astronomy, Inc., under NASA contract
NAS5-26555. Support for MAST for non-HST data is provided by the NASA Of-
fice of Space Science via grant NNX09AF08G and by other grants and contracts.





\bibliographystyle{mnras}
\bibliography{szewczuk}


\appendix
\section{Frequencies of KIC\,3240411}
\label{app_freq}

Table\,\ref{tab_freq} contains all frequencies extracted from the {\it Kepler} data for KIC\,3240411.
The number of frequency peaks with $S/N\ge 4$ is 389 and with $S/N\ge 5$ it decreases to 72.
Thus the vast majority of the frequencies has the $S/N$ ratio between 4 and 5.
There are frequencies whose differences are lower than the Rayleigh resolution (see the last column of Table\,\ref{tab_freq}).
The Rayleigh resolution of the full data set in the case of KIC\,3240411 is equal to 0.00068.
Besides, a vast majority of frequencies satisfy a mathematical condition for combinations. However, due to the density of the extracted oscillation spectrum
most of them are rather independent (see main text).

In Table\,\ref{tab_freq_small_stars} we give frequencies found in two sources identified in the {\it Kepler} frames
which are close to the target star.

\begin{table*}
	\centering
	\caption{Parameters of all frequencies extracted from the Kepler light curve of KIC\,3240411.
	There are given frequencies, amplitudes, phases and their errors, respectively.
	In the third column from the end are listed signal to noise ratios.
	In the penultimate column are listed possible combinations. If a given frequency is closer to the other with higher amplitude than the Rayleigh resolution,
    then this higher amplitude frequency is listed  in the last column.}
	\label{tab_freq}
\begin{tabular}{rrrrrrrrrr}
	\hline
    ID&    $\nu$ [d$^{-1}$] & A [ppt] & $\phi$ [0--1] & $\epsilon_\nu$ [d$^{-1}$] & $\epsilon_\mathrm{A}$ [ppt] & $\epsilon_\phi$ [0--1] & S/N & combinations & remarks\\
	\hline
    1 &    0.915583 &      0.5981 &      0.5649 &    0.000003 &      0.0019 &      0.0005 &   36.9 &   &  \\
    2 &    0.914934 &      0.5784 &      0.4247 &    0.000004 &      0.0019 &      0.0005 &   38.4 &   & $\approx \nu_1$ \\
    3 &    0.721591 &      0.3582 &      0.1980 &    0.000004 &      0.0017 &      0.0007 &   27.5 &   &  \\
    4 &    1.453060 &      0.3579 &      0.6010 &    0.000004 &      0.0017 &      0.0007 &   37.5 &   &  \\
    5 &    1.193358 &      0.3612 &      0.5842 &    0.000004 &      0.0017 &      0.0007 &   30.6 &   &  \\
    6 &    0.119755 &      0.2399 &      0.7654 &    0.000006 &      0.0017 &      0.0011 &   21.6 &   &  \\
    7 &    1.195610 &      0.2402 &      0.6084 &    0.000006 &      0.0023 &      0.0015 &   21.4 &   &  \\
\hline
\end{tabular}
\end{table*}

\begin{table*}
	\centering
	\contcaption{}
	\label{tab_freq}
	\begin{tabular}{rrrrrrrrrr}
	\hline
    ID&    $\nu$ [d$^{-1}$] & A [ppt] & $\phi$ [0--1] & $\epsilon_\nu$ [d$^{-1}$] & $\epsilon_\mathrm{A}$ [ppt] & $\epsilon_\phi$ [0--1] & S/N & combinations & remarks\\
	\hline

   61 &    0.509253 &      0.0394 &      0.1670 &    0.000020 &      0.0017 &      0.0067 &    4.8 &   &  \\
   62 &    0.064882 &      0.0400 &      0.0072 &    0.000020 &      0.0017 &      0.0066 &    5.1 & 2$\nu_{3}$-9$\nu_{57}$  &  \\
   63 &    1.128600 &      0.0391 &      0.5913 &    0.000020 &      0.0017 &      0.0068 &    5.2 & -7$\nu_{39}$+8$\nu_{57}$  &  \\
   64 &    1.237658 &      0.0413 &      0.5979 &    0.000028 &      0.0018 &      0.0068 &    5.9 &   &  $\approx \nu_9$\\
   65 &    0.673129 &      0.0356 &      0.3875 &    0.000020 &      0.0017 &      0.0075 &    4.9 &   &  \\
   66 &    0.828615 &      0.0631 &      0.8380 &    0.000035 &      0.0024 &      0.0060 &    4.9 &   & $\approx \nu_{23}$ \\
   67 &    0.558367 &      0.0390 &      0.3265 &    0.000020 &      0.0017 &      0.0068 &    4.9 & 5$\nu_{25}$-6$\nu_{56}$  &  \\
\hline
\end{tabular}
\end{table*}

\begin{table*}
	\centering
	\contcaption{}
	\label{tab_freq}
	\begin{tabular}{rrrrrrrrrr}
	\hline
    ID&    $\nu$ [d$^{-1}$] & A [ppt] & $\phi$ [0--1] & $\epsilon_\nu$ [d$^{-1}$] & $\epsilon_\mathrm{A}$ [ppt] & $\epsilon_\phi$ [0--1] & S/N & combinations & remarks\\
	\hline

  121 &    1.210427 &      0.0267 &      0.7625 &    0.000026 &      0.0017 &      0.0100 &    4.4 & -5$\nu_{23}$+7$\nu_{25}$  &  \\
  122 &    1.344880 &      0.0279 &      0.9753 &    0.000026 &      0.0017 &      0.0097 &    4.7 & 6$\nu_{12}$-5$\nu_{23}$  &  \\
  123 &    1.195939 &      0.0385 &      0.5692 &    0.000045 &      0.0022 &      0.0093 &    4.2 &   &  $\approx \nu_7$\\
  124 &    1.194645 &      0.0254 &      0.4013 &    0.000030 &      0.0017 &      0.0108 &    4.2 & 4$\nu_{7}$-3$\nu_{123}$  &  \\
  125 &    1.253989 &      0.0243 &      0.1763 &    0.000028 &      0.0017 &      0.0110 &    4.3 & 8$\nu_{19}$-7$\nu_{66}$  &  \\
  126 &    2.585108 &      0.0236 &      0.1088 &    0.000028 &      0.0017 &      0.0112 &    9.3 & -2$\nu_{7}$+7$\nu_{37}$  &  \\
  127 &    1.484467 &      0.0234 &      0.5614 &    0.000029 &      0.0017 &      0.0113 &    4.7 & -$\nu_{72}$+3$\nu_{99}$  &  \\
\hline
\end{tabular}
\end{table*}

\begin{table*}
	\centering
	\contcaption{}
	\label{tab_freq}
	\begin{tabular}{rrrrrrrrrr}
	\hline
    ID&    $\nu$ [d$^{-1}$] & A [ppt] & $\phi$ [0--1] & $\epsilon_\nu$ [d$^{-1}$] & $\epsilon_\mathrm{A}$ [ppt] & $\epsilon_\phi$ [0--1] & S/N & combinations & remarks\\
	\hline

  181 &    2.223115 &      0.0116 &      0.6195 &    0.000055 &      0.0017 &      0.0228 &    4.5 & -3$\nu_{1}$+6$\nu_{23}$  &  \\
  182 &    3.133177 &      0.0117 &      0.6595 &    0.000055 &      0.0017 &      0.0226 &    4.9 & -3$\nu_{5}$+9$\nu_{22}$  &  \\
  183 &    3.824523 &      0.0115 &      0.8732 &    0.000056 &      0.0017 &      0.0230 &    4.8 & 3$\nu_{1}$+9$\nu_{6}$  &  \\
  184 &   10.690927 &      0.0112 &      0.6312 &    0.000057 &      0.0017 &      0.0237 &   17.5 & 3$\nu_{19}$+10$\nu_{33}$  &  \\
  185 &    3.274436 &      0.0112 &      0.9179 &    0.000058 &      0.0017 &      0.0237 &    4.7 & -9$\nu_{28}$+9$\nu_{46}$  &  \\
  186 &    2.132521 &      0.0478 &      0.5469 &    0.000101 &      0.0057 &      0.0190 &    4.1 & 7$\nu_{2}$-5$\nu_{18}$  & $\approx \nu_{27}$ \\
  187 &    1.759684 &      0.0140 &      0.3188 &    0.000046 &      0.0017 &      0.0189 &    4.0 & -9$\nu_{13}$+10$\nu_{19}$  &  \\
\hline
\end{tabular}
\end{table*}

\begin{table*}
	\centering
	\contcaption{}
	\label{tab_freq}
	\begin{tabular}{rrrrrrrrrr}
	\hline
    ID&    $\nu$ [d$^{-1}$] & A [ppt] & $\phi$ [0--1] & $\epsilon_\nu$ [d$^{-1}$] & $\epsilon_\mathrm{A}$ [ppt] & $\epsilon_\phi$ [0--1] & S/N & combinations & remarks\\
	\hline

  241 &   12.823955 &      0.0063 &      0.6613 &    0.000102 &      0.0017 &      0.0422 &   10.7 &   &  \\
  242 &   11.335626 &      0.0062 &      0.6306 &    0.000103 &      0.0017 &      0.0426 &   10.9 & $\nu_{10}$+7$\nu_{51}$  &  \\
  243 &   15.401603 &      0.0059 &      0.8077 &    0.000104 &      0.0017 &      0.0449 &   10.4 & 9$\nu_{10}$+6$\nu_{97}$  &  \\
  244 &   13.481200 &      0.0057 &      0.4105 &    0.000104 &      0.0017 &      0.0474 &    9.8 & 5$\nu_{39}$+7$\nu_{104}$  &  \\
  245 &   12.613991 &      0.0059 &      0.6593 &    0.000100 &      0.0017 &      0.0463 &   10.5 & -2$\nu_{36}$+$\nu_{224}$  &  \\
  246 &    8.070051 &      0.0057 &      0.0620 &    0.000110 &      0.0017 &      0.0460 &    5.2 & -2$\nu_{1}$+8$\nu_{64}$  &  \\
  247 &   13.995221 &      0.0052 &      0.8317 &    0.000126 &      0.0017 &      0.0531 &    8.5 & 6$\nu_{30}$+6$\nu_{101}$  &  \\
\hline
\end{tabular}
\end{table*}

\begin{table*}
	\centering
	\contcaption{}
	\label{tab_freq}
	\begin{tabular}{rrrrrrrrrr}
	\hline
    ID&    $\nu$ [d$^{-1}$] & A [ppt] & $\phi$ [0--1] & $\epsilon_\nu$ [d$^{-1}$] & $\epsilon_\mathrm{A}$ [ppt] & $\epsilon_\phi$ [0--1] & S/N & combinations & remarks\\
	\hline

  301 &   13.992423 &      0.0034 &      0.5672 &    0.000207 &      0.0017 &      0.0784 &    5.0 & $\nu_{2}$+9$\nu_{4}$  &  \\
  302 &    9.833278 &      0.0030 &      0.5167 &    0.000207 &      0.0017 &      0.0873 &    4.2 & 3$\nu_{7}$+9$\nu_{36}$  &  \\
  303 &   12.111014 &      0.0030 &      0.4882 &    0.000208 &      0.0017 &      0.0890 &    6.0 & 3$\nu_{6}$+9$\nu_{42}$  &  \\
  304 &   16.927602 &      0.0028 &      0.2916 &    0.000209 &      0.0017 &      0.0935 &    5.5 &   &  \\
  305 &   12.613181 &      0.0027 &      0.0258 &    0.000219 &      0.0017 &      0.1011 &    5.7 & 5$\nu_{12}$+7$\nu_{30}$  &  \\
  306 &   18.441031 &      0.0029 &      0.1125 &    0.000211 &      0.0017 &      0.0920 &    6.7 & 8$\nu_{9}$+10$\nu_{18}$  &  \\
  307 &   16.909717 &      0.0027 &      0.9921 &    0.000214 &      0.0017 &      0.0992 &    5.5 & 9$\nu_{4}$+2$\nu_{104}$  &  \\
\hline
\end{tabular}
\end{table*}

\begin{table*}
	\centering
	\contcaption{}
	\label{tab_freq}
	\begin{tabular}{rrrrrrrrrr}
	\hline
    ID&    $\nu$ [d$^{-1}$] & A [ppt] & $\phi$ [0--1] & $\epsilon_\nu$ [d$^{-1}$] & $\epsilon_\mathrm{A}$ [ppt] & $\epsilon_\phi$ [0--1] & S/N & combinations & remarks\\
	\hline

  361 &   12.987996 &      0.0023 &      0.3857 &    0.000277 &      0.0017 &      0.1139 &    4.5 & 4$\nu_{4}$+6$\nu_{123}$   &  \\
  362 &   14.004234 &      0.0023 &      0.1062 &    0.000301 &      0.0017 &      0.1140 &    4.1 & 9$\nu_{18}$+6$\nu_{47}$   &  \\
  363 &   12.999066 &      0.0023 &      0.3607 &    0.000275 &      0.0017 &      0.1146 &    4.4 & -2$\nu_{1}$+10$\nu_{51}$   &  \\
  364 &   14.241090 &      0.0023 &      0.2918 &    0.000283 &      0.0017 &      0.1165 &    4.0 & 3$\nu_{2}$+6$\nu_{104}$   &  \\
  365 &   12.675567 &      0.0041 &      0.8536 &    0.000507 &      0.0027 &      0.1073 &    4.5 & $\nu_{1}$+7$\nu_{148}$   & $\approx \nu_{209}$ \\
  366 &   15.669358 &      0.0022 &      0.4369 &    0.000285 &      0.0017 &      0.1189 &    4.2 & 7$\nu_{11}$+7$\nu_{99}$   &  \\
  367 &   11.193185 &      0.0022 &      0.3298 &    0.000283 &      0.0017 &      0.1192 &    4.0 & 2$\nu_{4}$+7$\nu_{101}$   &  \\
\hline
\end{tabular}
\end{table*}

\begin{table*}
	\centering
	\caption{The same as in Table.\,\ref{tab_freq} but for stars marked by magenta and cyan contours in Fig.\,2.}
	\label{tab_freq_small_stars}
	\begin{tabular}{rrrrrrrr}
	\hline
    ID&    $\nu$ [d$^{-1}$] & A [ppt] & $\phi$ [0--1] & $\epsilon_\nu$ [d$^{-1}$] & $\epsilon_\mathrm{A}$ [ppt] & $\epsilon_\phi$ [0--1] & S/N \\
	\hline
	\multicolumn{8}{c} {magenta mask star}\\
	\hline
    1 &   0.0207 &  0.214 &   0.96 &        0.00006 &        0.015 &        0.04 & 15.6 \\
    2 &   0.04902 &  0.176 &   0.05 &        0.00006 &        0.012 &        0.04 & 11.6 \\
    3 &   0.02980 &  0.127 &   0.19 &        0.00008 &        0.012 &        0.05 & 9.6 \\
    4 &   0.01580 &  0.122 &   0.72 &        0.00014 &        0.013 &        0.09 & 6.7 \\
    5 &   0.06078 &  0.086 &   0.16 &        0.00011 &        0.011 &        0.08 & 5.7 \\
    6 &   0.04242 &  0.091 &   0.99 &        0.00012 &        0.013 &        0.08 & 5.3 \\
    7 &   0.10425 &  0.078 &   0.61 &        0.00011 &        0.011 &        0.08 & 5.4 \\
\hline
	\multicolumn{8}{c} {cyan mask star}\\
	\hline
1 &	0.03189  &  0.164  &  0.14    &     0.00007    &     0.030     &    0.07 & 5.2 \\
	\hline
\end{tabular}
\end{table*}

\section{Known SPB stars with period spacing}

\begin{figure*}
	\includegraphics[width=1.4\columnwidth, angle=270]{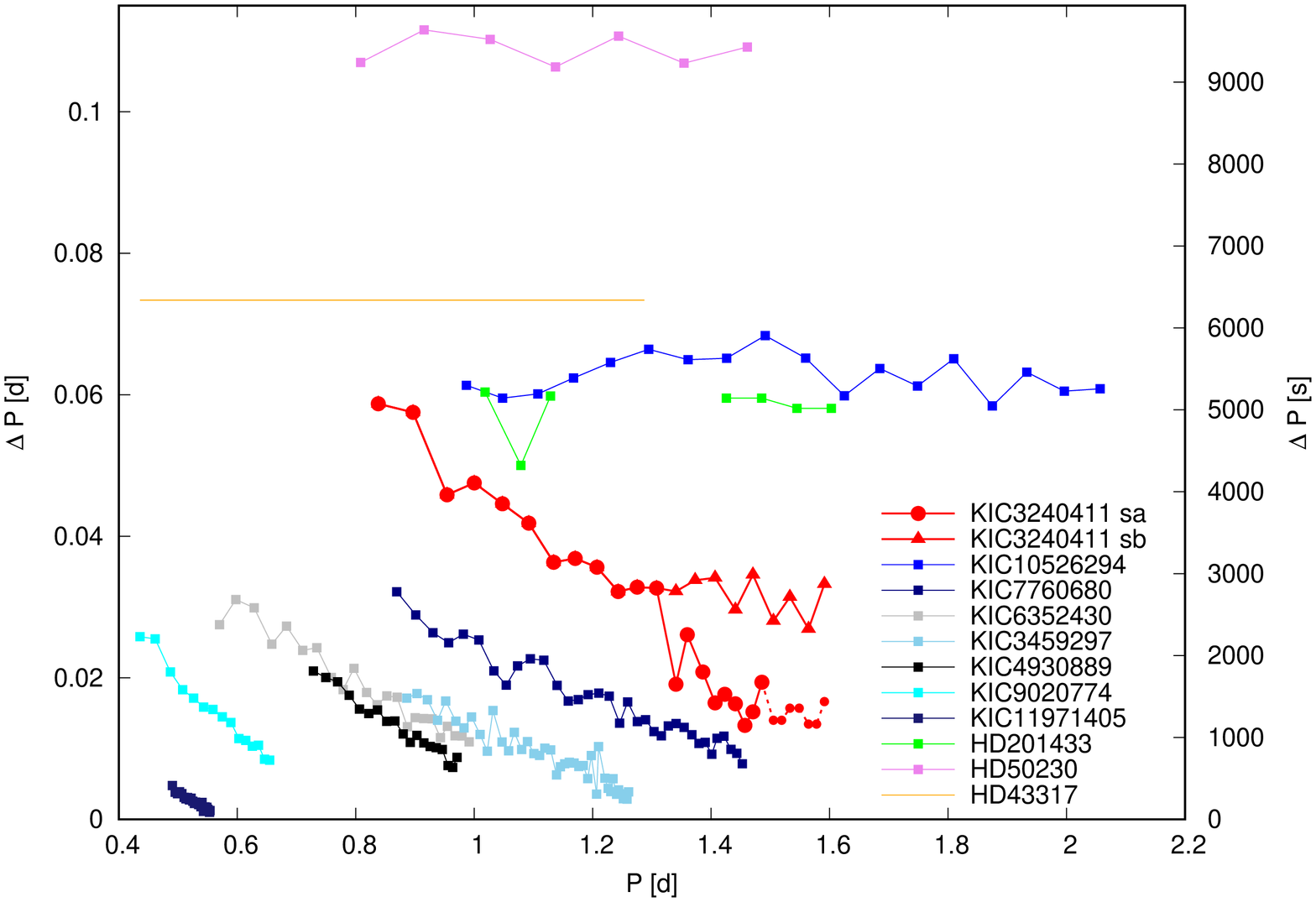}
    \caption{The period spacing as a function of period for all known B-type main sequence stars showing these structures. In the case of HD\,43317
             there is marked the mean period spacing because only this information was given in \citet{Papics2012}.}
    \label{per_spacings_8stars_eps}
\end{figure*}

So far, ten stars with period spacing patterns in their oscillation spectra have been detected.
Although in the two cases (HD\,50230 and HD\,43317) the connection of such patterns
with asymptotic properties is doubtful we put all of them in the period spacing $vs.$ period diagram (Fig.\,\ref{per_spacings_8stars_eps}).
In Table\,\ref{stellar_all_apr} we give stellar parameters of these stars. In the case of multiple systems we listed also parameters of companions if known.

\begin{table*}
	\centering
	\caption{Stellar parameters of stars shown in Fig.\,1 and \ref{per_spacings_8stars_eps} and the source of observations in which period patterns
	were found. The references are given in the last column.}
	\label{stellar_all_apr}
	\begin{tabular}{ccccccccc}
	\hline
    ID&  SpT  & $\log T_\mathrm{eff}$ & $\log g$ & $V_\mathrm{rot} \sin i$ [km\,s$^{-1}$] & binarity & observations  & references  \\
	
	\hline
    HD\,43317             & B4IV-V & 4.225$\pm$0.003  &  3.9$\pm$0.1  &  106$\pm$4         &                                  &   {\it CoRoT}               & {\cite{Papics2012}} \\
    HD\,50230$_\mathrm A$ &  B3    &  4.267$\pm$0.023 &   $3.8\pm0.3$ &        $6.9\pm1.5$ &   \multirow{2}{*}{  \checkmark } & \multirow{2}{*}{\it CoRoT}  & \multirow{2}{*} {\cite{Degroote2010N,2012A&A...542A..88D}}    \\
    HD\,50230$_\mathrm B$ &        &  $\le 16000$     &               &        117         &                                  &      \\
    HD\,201433            & B9V    & 4.077$\pm$0.007  & 4.15$\pm$0.07 &  9.8$\pm$2.2       &  \checkmark (triple)             &   {\it BRITE}               &  {\cite{2017A&A...603A..13K}}\\
{\bf KIC\,3240411}        & B2V    & 4.322$\pm$0.018  & 4.01$\pm$0.12 &  43$\pm$5          &                                  &  {\it Kepler}               & {\cite{Lehmann2011}}\\
 KIC\,3459297             & B6.5IV & 4.128$\pm$0.008  &  3.8$\pm$0.1  & 109$\pm$14         &                                  &  {\it Kepler}               & {\cite{Papics2017}} \\
 KIC\,4930889$_\mathrm A$ & B5IV-V & 4.179$\pm$0.004  & 3.95$\pm$0.1  & 116$\pm$6          &  \multirow{2}{*}{  \checkmark }  &  \multirow{2}{*}{\it Kepler}& \multirow{2}{*} {\cite{Papics2017}}    \\
 KIC\,4930889$_\mathrm B$ & B8IV-V & 4.082$\pm$0.007  & 3.85$\pm$0.1  & 85$\pm$5           &                                  &                             & \\
 KIC\,6352430$_\mathrm A$ & B7V    & 4.108$\pm$0.007  & 4.05$\pm$0.05 & 69.8$\pm$2.0       &  \multirow{2}{*}{  \checkmark }  &  \multirow{2}{*}{\it Kepler}& \multirow{2}{*} {\cite{2013A&A...553A.127P}}    \\
 KIC\,6352430$_\mathrm B$ & F2.5V  & 3.833$\pm$0.006  & 4.26$\pm$0.15 & 9.8$\pm$1.0        &                                  &                             & \\
 KIC\,7760680             & B8V    & 4.066$\pm$0.008  & 3.97$\pm$0.08 &  62$\pm$5          &                                  &  {\it Kepler}               & {\cite{Papics2015}} \\
 KIC\,9020774             & B5.7V  & 4.158$\pm$0.020  & 4.25$\pm$0.18 & 129$\pm$28         &                                  &  {\it Kepler}               & {\cite{Papics2017}} \\
KIC\,10526294             & B8.3V  & 4.063$\pm$0.019  & 4.1$\pm$0.2   &                    &                                  &  {\it Kepler}               & {\cite{Papics2014}} \\
KIC\,11971405             & B5IV-Ve& 4.179$\pm$0.006  & 3.94$\pm$0.06 & 242$\pm$14         &                                  &  {\it Kepler}               & {\cite{Papics2017}} \\
\hline
\end{tabular}
\end{table*}

\section{Correlations}
\label{app_cor}
\begin{figure*}
	\includegraphics[width=1.9\columnwidth, angle=270]{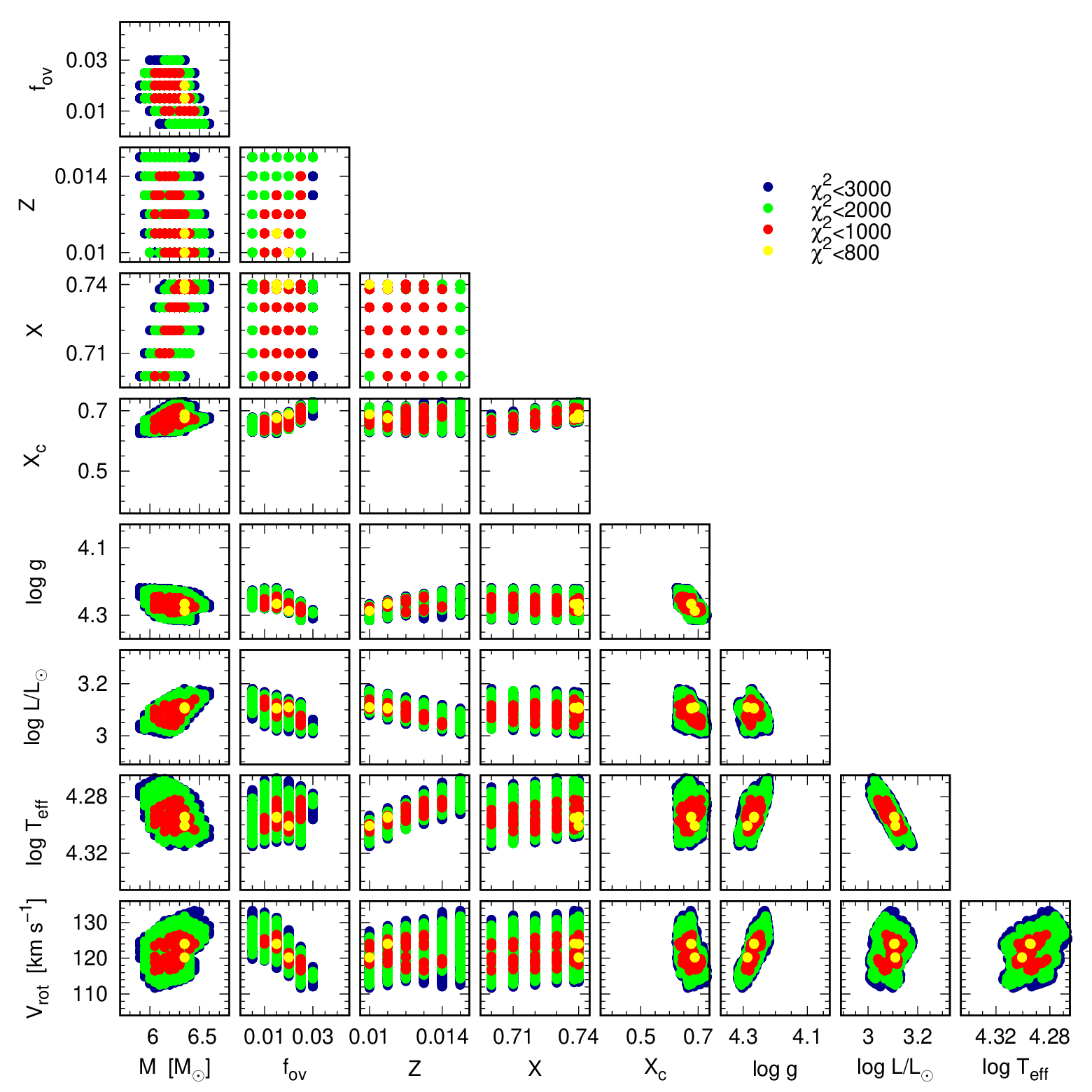}
    \caption{The correlations between different parameters in the vicinity of the seismic model \#19}
    \label{chi_OPLIB_set6_korel_size_reduced}
\end{figure*}

We also checked correlations between various parameters in our seismic modelling.
To this end we choose the vicinity of seismic model \#19, i.e. grid of models from which we found model \#22. 
In Fig.\,\ref{chi_OPLIB_set6_korel_size_reduced} we put the values of $\chi^2$ coded by colours in the parameter $vs.$ parameter diagram.
The correlation or anty-correlation occur between different parameters. For example, one can see the correlation between $V_\mathrm{rot}$ and $\log g$,
and anty-correlation between $V_\mathrm{rot}$ and the overshooting parameter $f_\mathrm{ov}$.
The metallicity correlates with the effective temperature and luminosity but we did not find the correlation between metallicity and other parameters
We got the clear correlation between $\log T_{\rm eff}$ and $\log L/\log \mathrm{L}_{\sun}$ what is easily  understood.
As one can expect, the correlation occurs also between the central value of hydrogen and the overshooting parameter; the higher value of $f_\mathrm{ov}$
the higher $X_c$.

\section{MESA inlist}

\texttt{\noindent \&star\_job}
\newline
\newline
\texttt{
history\_columns\_file = 'history\_columns1.list'\\
profile\_columns\_file = 'profile\_columns2.list'
}
\newline
\newline
\texttt{
set\_to\_this\_tau\_factor = 0.0001\\
set\_tau\_factor = .true.
}
\newline
\newline
\texttt{
change\_lnPgas\_flag = .true.\\
new\_lnPgas\_flag = .true.
}
\newline
\newline
\texttt{
new\_rotation\_flag = .true.\\
change\_rotation\_flag = .true.\\
set\_surf\_rotation\_v\_step\_limit = 1 !(controlled in run\_star\_extras.f)\\
new\_surface\_rotation\_v = var ! (km sec$^-1$)\\
set\_surface\_rotation\_v = .true.
}
\newline
\newline
\texttt{
change\_net = .true.\\
new\_net\_name = 'approx21.net'
}
\newline
\newline
\texttt{
set\_rates\_preference = .true.\\
new\_rates\_preference = 1\\
set\_rate\_c12ag = 'Kunz'\\
set\_rate\_n14pg = 'jina reaclib'\\
set\_rate\_3a = 'jina reaclib'\\
}
\newline
\newline
\texttt{
kappa\_file\_prefix = 'var'\\
kappa\_lowT\_prefix = 'lowT\_fa05\_a09p'
}
\newline
\newline
\texttt{/ !end of star\_job namelist}
\newline
\newline
\newline
\newline
\texttt{\&controls}
\newline
\newline
\texttt{
initial\_mass = var\\
initial\_z = var\\
initial\_y = var
}
\newline
\newline
\texttt{
terminal\_interval = 50\\
history\_interval = 1\\
profile\_interval = 1\\
max\_num\_profile\_models = 1000\\
photo\_interval = 5000
}  
\newline
\newline
\texttt{
xa\_central\_lower\_limit\_species(1) = 'h1'\\
xa\_central\_lower\_limit(1) = 0.01
}
\newline
\newline
\texttt{
which\_atm\_option = 'Eddington\_grey'
}  
\newline
\newline
\texttt{
do\_element\_diffusion = .true.\\
diffusion\_dt\_limit = 3.15e13     !MIST\\
diffusion\_min\_T\_at\_surface = 1d3 !MIST\\
diffusion\_num\_classes = 8 
diffusion\_class\_representative(1) = 'h1'\\
diffusion\_class\_representative(2) = 'he3'\\
diffusion\_class\_representative(3) = 'he4'\\
diffusion\_class\_representative(4) = 'c12'\\
diffusion\_class\_representative(5) = 'n14'\\
diffusion\_class\_representative(6) = 'o16'\\
diffusion\_class\_representative(7) = 'ne20'\\
diffusion\_class\_representative(8) = 'fe56'\\
}
\newline
\texttt{
! in ascending order.  species goes into 1st class with A\_max >= species A\\
diffusion\_class\_A\_max(1) = 2\\
diffusion\_class\_A\_max(2) = 3\\
diffusion\_class\_A\_max(3) = 4\\
diffusion\_class\_A\_max(4) = 12\\
diffusion\_class\_A\_max(5) = 14\\
diffusion\_class\_A\_max(6) = 16\\
diffusion\_class\_A\_max(7) = 20\\
diffusion\_class\_A\_max(8) = 10000\\
}    
\newline
\newline
\texttt{radiation\_turbulence\_coeff = 1.0 !MIST}
\newline
\newline
\texttt{
!CORE MASS DEFINITION\\
he\_core\_boundary\_h1\_fraction = 1d-4\\
c\_core\_boundary\_he4\_fraction = 1d-4\\
o\_core\_boundary\_c12\_fraction = 1d-4
}
\newline
\newline
\texttt{MLT\_option = 'Henyey'}
\newline
\newline
\texttt{
gradT\_excess\_f2 = 0.01\\
gradT\_excess\_max\_logT = 6.5
}
\newline
\newline
\texttt{
am\_D\_mix\_factor = 0.0228  ! Brott et al. (2011)\\
am\_gradmu\_factor = 0.1 ! Yoon et al (2006)
}
\newline
\newline
\texttt{
set\_uniform\_am\_nu\_non\_rot = .true.\\
uniform\_am\_nu\_non\_rot = 1d20  
}
\newline
\newline
\texttt{
D\_DSI\_factor = 1\\
D\_SH\_factor = 1\\
D\_SSI\_factor = 1\\
D\_ES\_factor = 1\\
D\_GSF\_factor = 1\\
D\_ST\_factor = 0
}
\newline
\newline
\texttt{cubic\_interpolation\_in\_Z = .true.}
\newline
\newline
\texttt{mixing\_length\_alpha = 1.5}
\newline
\newline
\texttt{
use\_Ledoux\_criterion = .true.\\
alpha\_semiconvection = 0.01
}
\newline
\newline
\texttt{
overshoot\_f\_above\_nonburn\_core = 0.02\\
overshoot\_f0\_above\_nonburn\_core = 0.02\\
overshoot\_f\_below\_nonburn\_shell = 0.02\\
overshoot\_f0\_below\_nonburn\_shell = 0.02\\
overshoot\_f\_above\_burn\_h\_core = var\\
overshoot\_f0\_above\_burn\_h\_core = 0.02\\
overshoot\_f\_above\_burn\_h\_shell = 0.02\\
overshoot\_f0\_above\_burn\_h\_shell = 0.02\\
overshoot\_f\_above\_burn\_he\_core = 0.02\\
overshoot\_f0\_above\_burn\_he\_core = 0.02
D\_mix\_ov\_limit = 1d0
}
\newline
\newline
\texttt{
!MESH AND TIMESTEP PARAMETERS\\
mesh\_delta\_coeff = 0.5\\
varcontrol\_target = 1d-4\\
max\_allowed\_nz = 50000\\
delta\_lg\_XH\_cntr\_max = -1\\
delta\_lgTeff\_limit = 0.00066\\
delta\_lgTeff\_hard\_limit = 0.01\\
delta\_lgL\_limit = 0.02\\
delta\_lgL\_hard\_limit = 0.05\\
mesh\_logX\_species(1) = 'h1'\\
mesh\_dlogX\_dlogP\_extra(1) = 0.25\\
mesh\_logX\_species(2) = 'he4'\\
mesh\_dlogX\_dlogP\_extra(2) = 0.25\\
mesh\_logX\_species(3) = 'c12'\\
mesh\_dlogX\_dlogP\_extra(3) = 0.25\\
mesh\_dlog\_3alf\_dlogP\_extra = 0.15\\
mesh\_dlog\_burn\_c\_dlogP\_extra = 0.15\\
mesh\_dlog\_burn\_n\_dlogP\_extra = 0.15\\
mesh\_dlog\_burn\_o\_dlogP\_extra = 0.15\\
mesh\_dlog\_burn\_ne\_dlogP\_extra = 0.15\\
mesh\_dlog\_burn\_na\_dlogP\_extra = 0.15\\
mesh\_dlog\_burn\_mg\_dlogP\_extra = 0.15\\
mesh\_dlog\_cc\_dlogP\_extra = 0.15\\
mesh\_dlog\_co\_dlogP\_extra = 0.15\\
mesh\_dlog\_oo\_dlogP\_extra = 0.15\\
mesh\_dlog\_burn\_si\_dlogP\_extra = 0.15\\
mesh\_dlog\_burn\_s\_dlogP\_extra = 0.15\\
mesh\_dlog\_burn\_ar\_dlogP\_extra = 0.15\\
mesh\_dlog\_burn\_ca\_dlogP\_extra = 0.15\\
mesh\_dlog\_burn\_ti\_dlogP\_extra = 0.15\\
mesh\_dlog\_burn\_cr\_dlogP\_extra = 0.15\\
mesh\_dlog\_burn\_fe\_dlogP\_extra = 0.15\\
mesh\_dlog\_pnhe4\_dlogP\_extra = 0.15\\
mesh\_dlog\_other\_dlogP\_extra = 0.15\\
\newline
xtra\_coef\_os\_above\_nonburn = 0.25\\
xtra\_coef\_os\_below\_nonburn = 0.25\\
xtra\_coef\_os\_above\_burn\_h = 0.25\\
xtra\_coef\_os\_below\_burn\_h = 0.25\\
xtra\_coef\_os\_above\_burn\_he = 0.25\\
xtra\_coef\_os\_below\_burn\_he = 0.25\\
xtra\_coef\_os\_above\_burn\_z = 0.25\\
xtra\_coef\_os\_below\_burn\_z = 0.25
}      
\newline
\newline
\texttt{zams\_filename = 'var'}
\newline
\newline
\texttt{/ ! end of controls namelist}
\newline
\newline
\texttt{\&pgstar}
\newline
\newline
\texttt{/ ! end of pgstar namelist}

\bsp	
\label{lastpage}
\end{document}